\documentclass[useAMS,usenatbib,a4paper]{mn2e}
\usepackage{amsmath,amsfonts,latexsym,graphicx,amssymb,times,appendix,subfigure}
\newcommand{\bsf}[1]{\textbf{\textsf{#1}}}

\title[BAMBI]{BAMBI:  blind accelerated multimodal Bayesian inference}
\author[Graff, Feroz, Hobson, and Lasenby]{Philip Graff$^1$\thanks{Email: \texttt{p.graff@mrao.cam.ac.uk}}, Farhan Feroz $^1$, Michael P. Hobson$^1$, and Anthony Lasenby$^{1,2}$\\
		$^1$Astrophysics Group, Cavendish Laboratory, JJ Thomson Avenue, Cambridge CB3 0HE, UK\\
		$^2$Kavli Institute for Cosmology, Madingley Road, Cambridge CB3 0HA, UK}
\date{\today}

\pagerange{\pageref{firstpage}--\pageref{lastpage}} \pubyear{2011}

\begin{document}

\label{firstpage}

\maketitle

\begin{abstract}
In this paper, we present an algorithm for rapid Bayesian analysis that combines the benefits of nested sampling and artificial neural networks. The blind accelerated multimodal Bayesian inference (BAMBI) algorithm implements the {\sc MultiNest} package for nested sampling as well as the training of an artificial neural network (NN) to learn the likelihood function. In the case of computationally expensive likelihoods, this allows the substitution of a much more rapid approximation in order to increase significantly the speed of the analysis. We begin by demonstrating, with a few toy examples, the ability of a NN to learn complicated likelihood surfaces. BAMBI's ability to decrease running time for Bayesian inference is then demonstrated in the context of estimating cosmological parameters from {\em Wilkinson Microwave Anisotropy Probe} and other observations. We show that valuable speed increases are achieved in addition to obtaining NNs trained on the likelihood functions for the different model and data combinations. These NNs can then be used for an even faster follow-up analysis using the same likelihood and different priors. This is a fully general algorithm that can be applied, without any pre-processing, to other problems with computationally expensive likelihood functions.
\end{abstract}


\section{Introduction}\label{sec:intro}
Bayesian methods of inference are widely used in astronomy and cosmology and are gaining popularity in other fields, such as particle physics. They can generally be divided into the performance of two main tasks:  parameter estimation and model selection. The former has traditionally been performed using Markov chain Monte Carlo (MCMC) methods, usually based on the Metropolis-Hastings algorithm or one of its variants. These can be computationally expensive in their exploration of the parameter space and often need to be finely tuned in order to produce accurate results. Additionally, sampling efficiency can be seriously affected by multimodal distributions and large, curving degeneracies. The second task of model selection is further hampered by the need to calculate the Bayesian evidence accurately. The most common method of doing so is thermodynamic integration, which requires several chains to be run, thus multiplying the computational expense. Fast methods of evidence calculation, such as assuming a Gaussian peak, clearly fail in multimodal and degenerate situations. Nested sampling~\citep{SkillingNest} is a method of Monte Carlo sampling designed for efficient calculation of the evidence which also provides samples from the posterior distribution as a by-product, thus allowing parameter estimation at no additional cost. The {\sc MultiNest} algorithm~\citep{MultiNest1,MultiNest2} is a generic implementation of nested sampling, extended to handle multimodal and degenerate distributions, and is fully parallelised.

At each point in parameter space, Bayesian methods require the evaluation of a `likelihood' function describing the probability of obtaining the data for a given set of model parameters. For some cosmological and particle physics problems each such function evaluation takes up to tens of seconds. MCMC applications may require millions of these evaluations, making them prohibitively costly. {\sc MultiNest} is able to reduce the number of likelihood function calls by an order of magnitude or more, but further gains can be achieved if we are able to speed up the evaluation of the likelihood itself. An artificial neural network (NN) is ideally suited for this task. A universal approximation theorem assures us that we can accurately and precisely approximate the likelihood with a NN of a given form. The training of NNs is one of the most widely studied problems in machine learning, so techniques for learning the likelihood function are well established. We implement a variant of conjugate gradient descent to find the optimum set of weights for a NN, using regularisation of the likelihood and a Hessian-free second-order approximation to improve the quality of proposed steps towards the best fit.

The blind accelerated multimodal Bayesian inference (BAMBI) algorithm combines these two elements. After a specified number of new samples from {\sc MultiNest} have been obtained, BAMBI uses these to train a network on the likelihood function. After convergence to the optimal weights, we test that the network is able to predict likelihood values to within a specified tolerance level. If not, sampling continues using the original likelihood until enough new samples have been made for training to be done again. Once a network is trained that is sufficiently accurate, its predictions are used in place of the original likelihood function for future samples for {\sc MultiNest}. Using the network reduces the likelihood evaluation time from seconds to milliseconds, allowing {\sc MultiNest} to complete the analysis much more rapidly. As a bonus, the user also obtains a network that is trained to easily and quickly provide more likelihood evaluations near the peak if needed, or in subsequent analyses.

The structure of the paper is as follows. In Section~\ref{sec:bayesnest} we will introduce Bayesian inference and the use of nested sampling. Section~\ref{sec:NNoptimiser} will then explain the structure of a NN and how our optimiser works to find the best set of weights. We present some toy examples with BAMBI in Section~\ref{sec:toys} to demonstrate its capabilities; we apply BAMBI to cosmological parameter estimation in Section~\ref{sec:cosmology}. In Section~\ref{sec:NNanalysis} we show the full potential speed-up from BAMBI, by using the trained NNs in a follow-up analysis. Section~\ref{sec:conclusion} summarises our work and presents our conclusions.

\section{Bayesian Inference and {\sc MultiNest}}\label{sec:bayesnest}
\subsection{Theory of Bayesian Inference}\label{sec:inference}
Bayesian statistical methods provide a consistent way of estimating the probability distribution of a set of parameters $\mathbf{\Theta}$ for a given model or hypothesis $H$ given a data set $\mathbf{D}$. Bayes' theorem states that
\begin{equation}
\label{eq:bayes}
\Pr(\mathbf{\Theta}|\mathbf{D}, H) = \frac{\Pr(\mathbf{D}|\mathbf{\Theta},H)\Pr(\mathbf{\Theta}|H)}{\Pr(\mathbf{D}|H)},
\end{equation}
where $\Pr(\mathbf{\Theta}|\mathbf{D}, H)$ is the posterior probability distribution of the parameters and is written as $P(\mathbf{\Theta})$, $\Pr(\mathbf{D}|\mathbf{\Theta}, H)$ is the likelihood and is written as $\mathcal{L}(\mathbf{\Theta})$, $\Pr(\mathbf{\Theta}|H)$ is the prior distribution and is written as $\pi(\mathbf{\Theta})$, and $\Pr(\mathbf{D}|H)$ is the Bayesian evidence and is written as $\mathcal{Z}$. The evidence is the factor required to normalise the posterior over~$\mathbf{\Theta}$,
\begin{equation}
\label{eq:evidence}
\mathcal{Z} = \int{\mathcal{L}(\mathbf{\Theta})\pi(\mathbf{\Theta})}d^N\mathbf{\Theta},
\end{equation}
where $N$ is the dimensionality of the parameter space. Since the Bayesian evidence is independent of the parameter values, $\mathbf{\Theta}$, it can be ignored in parameter estimation problems and the posterior inferences obtained by exploring the un--normalized posterior.

Bayesian parameter estimation has achieved widespread use in many astrophysical applications. Standard Monte Carlo methods such as the Metropolis--Hastings algorithm or Hamiltonian sampling~\citep[see][]{MacKayIT} can experience problems with multimodal likelihood distributions, as they can get stuck in a single mode. Additionally, long and curving degeneracies are difficult for them to explore and can greatly reduce sampling efficiency. These methods often require careful tuning of proposal jump distributions and testing for convergence can be problematic. Additionally, calculation of the evidence for model selection often requires running multiple chains, greatly increasing the computational cost. Nested sampling and the {\sc Multinest} algorithm implementation address these problems.

\subsection{Nested Sampling}\label{sec:sampling}
\begin{figure}
\begin{center}
\subfigure[]{\includegraphics[width=0.4\columnwidth]{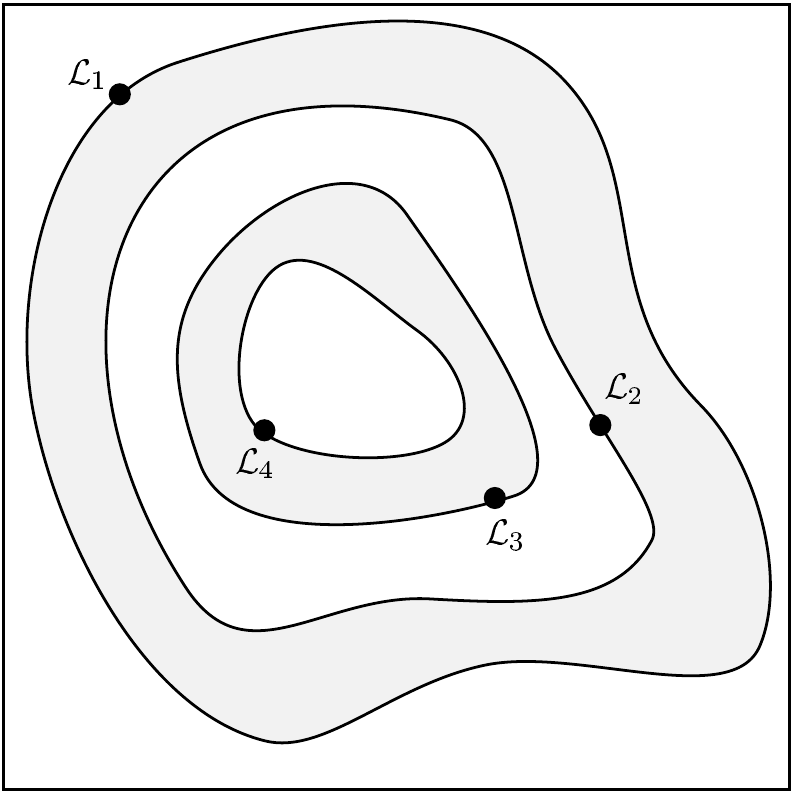}}\hspace{0.5cm}
\subfigure[]{\includegraphics[width=0.4\columnwidth]{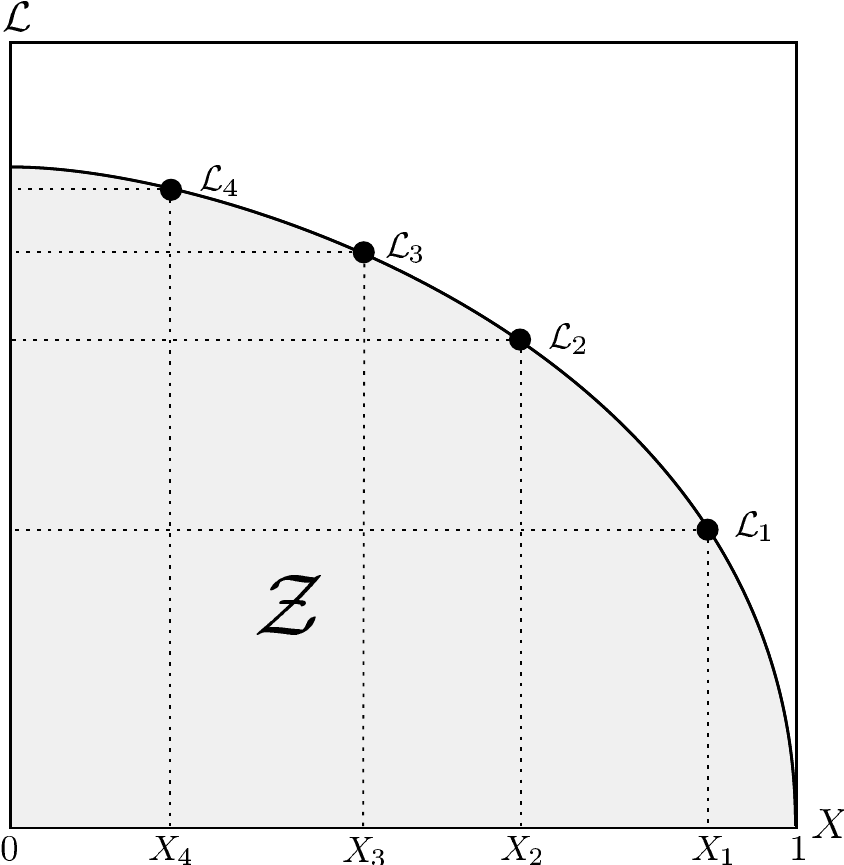}}
\caption{Cartoon illustrating (a) the posterior of a two dimensional problem; and (b) the transformed $\mathcal{L}(X)$  function where the prior volumes $X_{i}$ are associated with each likelihood $\mathcal{L}_{i}$. Originally published in Feroz \& Hobson (2008).}
\label{fig:NS}
\end{center}
\end{figure}

Nested sampling~\citep{SkillingNest} is a Monte Carlo method used for the computation of the evidence that can also provide posterior inferences. It transforms the multi-dimensional integral of Equation~\ref{eq:evidence} into a one-dimensional integral over the prior volume. This is done by defining the prior volume $X$ as $dX = \pi(\mathbf{\Theta})d^N \mathbf{\Theta}$. Therefore,
\begin{equation}
X(\lambda) = \int_{\mathcal{L}\left(\mathbf{\Theta}\right) > \lambda} \pi(\mathbf{\Theta}) d^N\mathbf{\Theta}.
\label{eq:Xdef}
\end{equation}
This integral extends over the region of parameter space contained within the likelihood contour $\mathcal{L}(\mathbf{\Theta}) = \lambda$. The evidence integral, Equation~\ref{eq:evidence}, can then be
written as
\begin{equation}
\label{eq:evidence2}
\mathcal{Z}=\int_{0}^{1}{\mathcal{L}(X)}dX,
\end{equation}
where $\mathcal{L}(X)$ is the inverse of Equation~\ref{eq:Xdef} and is a  monotonically decreasing function of $X$.  Thus, if we evaluate the likelihoods $\mathcal{L}_{i}=\mathcal{L}(X_{i})$, where $X_{i}$ is a sequence of decreasing values,
\begin{equation}
\label{eq:Xseq}
0<X_{M}<\cdots <X_{2}<X_{1}< X_{0}=1.
\end{equation}
The evidence can then be approximated numerically as a weighted sum
\begin{equation}
\label{eq:evidencesum}
\mathcal{Z}={\textstyle {\displaystyle \sum_{i=1}^{M}}\mathcal{L}_{i}w_{i}},
\end{equation}
where the weights $w_{i}$ for the simple trapezium rule are given by $w_{i}=\frac{1}{2}(X_{i-1}-X_{i+1})$. An example of a posterior in two dimensions and its associated function $\mathcal{L}(X)$ is shown in Figure~\ref{fig:NS}.

The fundamental operation of nested sampling begins with the initial, `live', points being chosen at random from the entire prior volume. The lowest likelihood live point is removed and replaced by a new sample with higher likelihood. This removal and replacement of live points continues until a stopping condition is reached ({\sc MultiNest} uses a tolerance on the evidence calculation). The difficult task lies in finding a new sample with higher likelihood than the discarded point. As the algorithm goes up in likelihood, the prior volume that will satisfy this condition decreases until it contains only a very small portion of the total parameter space, making this sampling potentially very inefficient. {\sc MultiNest} tackles this problem by enclosing all of the active points in clusters of ellipsoids. New points can then be chosen from within these ellipsoids using a fast analytic function. Since the ellipsoids will decrease in size along with the distribution of live points, their surfaces in effect represent likelihood contours of increasing value; the algorithm climbs up these contours seeking new points. As the clusters of ellipsoids are not constrained to fit any particular distribution, they can easily enclose curving degeneracies and are able to separate out to allow for multimodal distributions. This separation also allows for the calculation of the `local' evidence associated with each mode. {\sc MultiNest} has been shown to be of substantial use in astrophysics and particle physics~\citep[see][]{Feroz:2008wr,2008arXiv08100781F,2008arXiv08111199F,MNnospin,LISABurst,2008JHEP12024T,LISASMBH}, typically showing great improvement in efficiency over traditional MCMC techniques.

\section{Artificial Neural Networks and Weights Optimisation}\label{sec:NNoptimiser}
Artificial neural networks are a method of computation loosely based on the structure of a brain. They consist of a group of interconnected nodes, which process information that they receive and then pass this along to other nodes via weighted connections. We will consider only feed-forward NN, for which the structure is directed, with a layer of input nodes passing information to an output layer, via zero, one, or many hidden layers in between. A network is able `learn' a relationship between inputs and outputs given a set of training data and can then make predictions of the outputs for new input data. Further introduction can be found in~\cite{MacKayIT}.

\subsection{Network Structure}\label{sec:NNstruct}
\begin{figure}
\begin{center}
\includegraphics[width=2.5in]{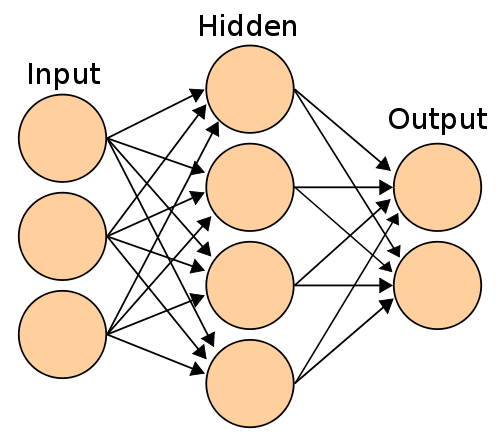}
\caption{A 3-layer neural network with 3 inputs, 4 hidden nodes, and 2 outputs. Image courtesy of Wikimedia Commons.}
\label{fig:neuralnet}
\end{center}
\end{figure}

A multilayer perceptron artificial neural network (NN) is the simplest type of network and consists of ordered layers of perceptron nodes that pass scalar values from one layer to the next. The perceptron is the simplest kind of node, and maps an input vector ${\bf x} \in \Re^n$ to a scalar output $f({\bf x};{\bf w},\theta)$ via
\begin{equation}
\label{eq:perceptron}
f({\bf x};{\bf w},\theta) = \theta + \sum_{i=1}^n {w_i x_i},
\end{equation}
where $\{w_i\}$ and $\theta$ are the parameters of the perceptron, called the `weights' and `bias', respectively. We will focus mainly on 3-layer NNs, which consist of an input layer, a hidden layer, and an output layer as shown in Figure~\ref{fig:neuralnet}.
The outputs of the nodes in the hidden and output layers are given by the following equations:
\begin{eqnarray}
\label{eq:hiddenformula}
\textrm{hidden layer:} \; h_j = g^{(1)}(f^{(1)}_j); \; f^{(1)}_j = \theta^{(1)}_j + \sum_l {w^{(1)}_{jl}x_l} ,\\
\label{eq:outputformula}
\textrm{output layer:} \; y_i = g^{(2)}(f^{(2)}_i); \; f^{(2)}_i = \theta^{(2)}_i + \sum_j {w^{(2)}_{ij}h_j} ,
\end{eqnarray}
where $l$ runs over input nodes, $j$ runs over hidden nodes, and $i$ runs over output nodes. The functions $g^{(1)}$ and $g^{(2)}$ are called activation functions and must be bounded, smooth, and monotonic for our purposes. We use $g^{(1)}(x)=\tanh(x)$ and $g^{(2)}(x)=x$; the non-linearity of $g^{(1)}$ is essential to allowing the network to model non-linear functions.

The weights and biases are the values we wish to determine in our training (described in Section~\ref{sec:NNtrain}). As they vary, a huge range of non-linear mappings from inputs to outputs is possible. In fact, a `universal approximation theorem'~\citep[see][]{UnivApprox} states that a NN with three or more layers can approximate any continuous function as long as the activation function is locally bounded, piecewise continuous, and not a polynomial (hence our use of $\tanh$, although other functions would work just as well, such as a sigmoid). By increasing the number of hidden nodes, we can achieve more accuracy at the risk of overfitting to our training data.

As long as the mapping from model parameters to predicted data is continuous -- and it is in many cases --  the likelihood function will also be continuous. This makes a NN an ideal tool for approximating the likelihood.

\subsection{Choosing the Number of Hidden Layer Nodes}\label{sec:nhid}
An important choice when using a NN is the number of hidden layer nodes to use. We consider here just the case of three-layer networks with only one hidden layer. The optimal number is a complex relationship between of the number of training data points, the number of inputs and outputs, and the complexity of the function to be trained. Choosing too few nodes will mean that the NN is unable to learn the likelihood function to sufficient accuracy. Choosing too many will increase the risk of overfitting to the training data and will also slow down the training process. As a general rule, a NN should not need more hidden nodes than the number of training data points. We choose to use $50$ or $100$ nodes in the hidden layer in our examples. These choices allow the network to model the complexity of the likelihood surface and its functional dependency on the input parameters. The toy examples (Section~\ref{sec:toys}) have fewer inputs that result in more complicated surfaces, but with simple functional relationships. The cosmological examples (Section~\ref{sec:cosmology}) have more inputs with complex relationships to generate a simple likelihood surface. In practice, it will be better to over-estimate the number of hidden nodes required. There are checks built in to prevent over-fitting and for computationally expensive likelihoods the additional training time will not be a large penalty if a usable network can be obtained earlier in the analysis.

\subsection{Network Training}\label{sec:NNtrain}
We wish to use a NN to model the likelihood function given some model and associated data. The input nodes are the model parameters and the single output is the value of the likelihood function at that point. The set of training data, $\mathcal{D} = \{{\bf x}^{(k)},t^{(k)}\}$, is the last \texttt{updInt} number of points accepted by {\sc MultiNest}, a parameter that is set by the user. This data is split into two groups randomly; approximately $80\%$ is used for training the network and $20\%$ is used as validation data to avoid overfitting. If training does not produce a sufficiently accurate network, {\sc MultiNest} will obtain $\texttt{updInt}/2$ new samples before attempting to train again. Additionally, the range of log-likelihood values must be within a user-specified range or NN training will be postponed.

\subsubsection{Overview}\label{sec:NNtrain_overview}
The weights and biases we will collectively call the network parameter vector ${\bf a}$. We can now consider the probability that a given set of network parameters is able to reproduce the known training data outputs -- representing how well our NN model of the original likelihood reproduces the true values. This gives us a log-likelihood function for ${\bf a}$, depending on a standard $\chi^2$ error function, given by
\begin{align}
\label{eq:chisquared}
\log(\mathcal{L}({\bf a};\sigma)) =& -\frac{K\log(2\pi)}{2}-\log(\sigma) \\ \notag
& -\frac{1}{2}\sum_{k=1}^K {\left[\frac{t^{(k)} - y({\bf x}^{(k)};{\bf a})}{\sigma}\right]^2},
\end{align}
where $K$ is the number of data points and $y({\bf x}^{(k)};{\bf a})$ is the NN's predicted output value for the inputs ${\bf x}^{(k)}$ and network parameters ${\bf a}$. The value of $\sigma$ is a hyper-parameter of the model that describes the standard deviation (error size) of the output. Our algorithm considers the parameters ${\bf a}$ to be probabilistic with a log-prior distribution given by
\begin{equation}
\label{eq:netprior}
\log(\mathcal{S}({\bf a};\alpha)) = -\frac{\alpha}{2} \sum_i {a_i^2}.
\end{equation}
$\alpha$ is a hyper-parameter of the model, called the `regularisation constant', that gives the relative influence of the prior and the likelihood. The posterior probability of a set of NN parameters is thus
\begin{equation}
\Pr({\bf a};\alpha,\sigma) \propto \mathcal{L}({\bf a};\sigma) \times \mathcal{S}({\bf a};\alpha).
\label{eq:netpost}
\end{equation}

BAMBI's network training begins by setting random values for the weights, sampled from a normal distribution with zero mean. The initial value of $\sigma$ is set by the user and can be set on either the true log-likelihood values themselves or on their whitened values (whitening involves performing a linear transform such that the training data values have a mean of zero and standard deviation of one). The only difference between these two settings is the magnitude of the error used. The algorithm then calculates a large initial estimate for $\alpha$,
\begin{equation}
\alpha = \frac{\lvert \nabla \log(\mathcal{L}) \rvert}{\sqrt{M r}},
\label{eq:initalpha}
\end{equation}
where $M$ is the total number of weights and biases (NN parameters) and $r$ is a rate set by the user ($0 < r \leq 1$, default $r=0.1$) that defines the size of the `confidence region' for the gradient. This formula for $\alpha$ sets larger regularisation (`damping') when the magnitude of the gradient of the likelihood is larger. This relates the amount of ``smoothing'' required to the steepness of the function being smoothed. The rate factor in the denominator allows us to increase the damping for smaller confidence regions on the value of the gradient. This results in smaller, more conservative steps that are more likely to result in an increase in the function value.

BAMBI then uses conjugate gradients to calculate a step, $\Delta {\bf a}$, that should be taken (see Section~\ref{sec:NNtrain_step}). Following a step, adjustments to $\alpha$ and $\sigma$ may be made before another step is calculated. The methods for calculating the initial $\alpha$ value and then determining subsequent adjustments of $\alpha$ and/or $\sigma$ are as developed for the {\sc MemSys} software package, described in~\cite{MemSys}.

\subsubsection{Finding the next step}\label{sec:NNtrain_step}
In order to find the most efficient path to an optimal set of parameters, we perform conjugate gradients using second-order derivative information. Newton's method gives the second-order approximation of a function,
\begin{equation}
f({\bf a}+\Delta {\bf a}) \approx f({\bf a}) + (\nabla f({\bf a}))^{\textrm{T}} \Delta {\bf a} + \frac{1}{2} (\Delta {\bf a})^{\textrm{T}} \bsf{B} \Delta {\bf a},
\label{eq:newton}
\end{equation}
where $\bsf{B}$ is the Hessian matrix of second derivatives of $f$ at ${\bf a}$. In this approximation, the maximum of $f$ will occur when
\begin{equation}
\nabla f({\bf a}+\Delta {\bf a}) \approx \nabla f({\bf a}) + \bsf{B} \Delta {\bf a} = 0.
\label{eq:newtongrad}
\end{equation}
Solving this for $\Delta {\bf a}$ gives us
\begin{equation}
\Delta {\bf a} = -\bsf{B}^{-1} \nabla f({\bf a}).
\label{eq:newtonsoln}
\end{equation}
Iterating this procedure will bring us eventually to the global maximum value of $f$. For our purposes, the function $f$ is the log-posterior distribution and hence Equation~\eqref{eq:newton} is a Gaussian approximation to the posterior. The Hessian of the log-posterior is the regularised (`damped') Hessian of the log-likelihood function, where the prior -- whose magnitude is set by $\alpha$ -- provides the regularisation. If we define the Hessian matrix of the log-likelihood as $\bsf{H}$, then $\bsf{B}=\bsf{H}+\alpha \bsf{I}$ ($\bsf{I}$ being the identity matrix).

Using the second-order information provided by the Hessian allows for more efficient steps to be made, since curvature information can extend step sizes in directions where the gradient varies less and shorten where it is varying more. Additionally, using the Hessian of the log-posterior instead of the log-likelihood adds the regularisation of the prior, which can help to prevent getting stuck in a local maximum by smoothing out the function being explored. It also aids in reducing the `region of confidence' for the gradient information which will make it less likely that a step results in a worse set of parameters.

Given the form of the log-likelihood, Equation~\eqref{eq:chisquared}, is a sum of squares (plus a constant), we can also save computational expense by utilising the Gauss-Newton approximation of its Hessian, given by
\begin{align}
\label{eq:gaussnewtonapprox}
\bsf{H}_{ij} =& -\sum_{k=1}^K \left( \frac{\partial r_k}{\partial a_i} \frac{\partial r_k}{\partial a_j} + r_k \frac{\partial^2 r_k}{\partial a_i \partial a_j} \right) \\ \notag
\approx& -\sum_{k=1}^K \left( \frac{\partial r_k}{\partial a_i} \frac{\partial r_k}{\partial a_j} \right),
\end{align}
where
\begin{equation}
r_k = \frac{t^{(k)} - y({\bf x}^{(k)};{\bf a})}{\sigma}.
\label{eq:gaussnewtonaddtl}
\end{equation}

The drawback of using second-order information is that calculation of the Hessian is computationally expensive and requires large storage, especially so in many dimensions as we will encounter for more complex networks. In general, the Hessian is not guaranteed to be positive semi-definite and so may not be invertible; however, the Gauss-Netwon approximation does have this guarantee. Inversion of the very large matrix will still be computationally expensive. As noted in~\cite{HessianFree} however, we only need products of the Hessian with a vector to compute the solution, never actually the full Hessian itself. To calculate these approximate Hessian-vector products, we use a fast approximate method given in~\cite{Schraudolph} and~\cite{Pearlmutter}. Combining all of these methods makes second-order information practical to use.

\subsubsection{Convergence}\label{sec:NNtrain_converge}
Following each step, the posterior, likelihood, and correlation values are calculated for the training data and the validation data that was not used in training (calculating the steps). Convergence to a best-fit set of parameters is determined by maximising the posterior, likelihood, or correlation of the validation data, as chosen by the user. This prevents overfitting as it provides a check that the network is still valid on points not in the training set. We use the correlation as the default function to maximise as it does not include the model hyper-parameters in its calculation.

\subsection{When to Use the Trained Network}\label{sec:netuse}
The optimal network possible with a given set of training data may not be able to predict likelihood values accurately enough, so an additional criterion is placed on when to use the trained network. This requirement is that the standard deviation of the difference between predicted and true log-likelihood values is less than a user-specified tolerance. When the trained network does not pass this test, then BAMBI will continue using the original log-likelihood function to obtain $\texttt{updInt}/2$ new samples to generate a new training data set of the last \texttt{updInt} accepted samples. Network training will then resume, beginning with the weights that it had found as optimal for the previous data set. Since samples are generated from nested contours and each new data set contains half of the previous one, the saved network will already be able to produce reasonable predictions on this new data; resuming therefore enables us to save time as fewer steps will be required to reach the new optimum weights.

Once a NN is in use in place of the original log-likelihood function, checks are made to ensure that the network is maintaining its accuracy. If the network makes a prediction outside of $[\textrm{min}(\textrm{training})-\sigma,\textrm{max}(\textrm{training})+\sigma]$, then that value is discarded and the original log-likelihood function is used for that point. Additionally, the central $95^{\textrm{th}}$ percentile of the output log-likelihood values from the training data used is calculated and if the network is making predictions mostly outside of this range then it will be re-trained. To re-train the network, BAMBI first substitutes the original log-likelihood function back in and gathers the required number of new samples from {\sc MultiNest}. Training then commences, resuming from the previously saved network. These criteria ensure that the network is not trusted too much when making predictions beyond the limits of the data it was trained on, as we cannot be sure that such predictions are accurate.

\section{BAMBI Toy Examples}\label{sec:toys}
In order to demonstrate the ability of BAMBI to learn and accurately explore multimodal and degenerate likelihood surfaces, we first tested the algorithm on a few toy examples. The eggbox likelihood has many separate peaks of equal likelihood, meaning that the network must be able to make predictions across many different areas of the prior. The Gaussian shells likelihood presents the problem of making predictions in a very narrow and curving region. Lastly, the Rosenbrock function gives a long, curving degeneracy that can also be extended to higher dimensions. They all require high accuracy and precision in order to reproduce the posterior truthfully and each presents unique challenges to the NN in learning the likelihood. It is important to note that running BAMBI on these problems required more time than the straightforward analysis; this was as expected since the actual likelihood functions are simple analytic functions that do not require much computational expense.

\subsection{Eggbox}\label{sec:eggbox}
\begin{figure}
\begin{center}
\includegraphics[width=3in]{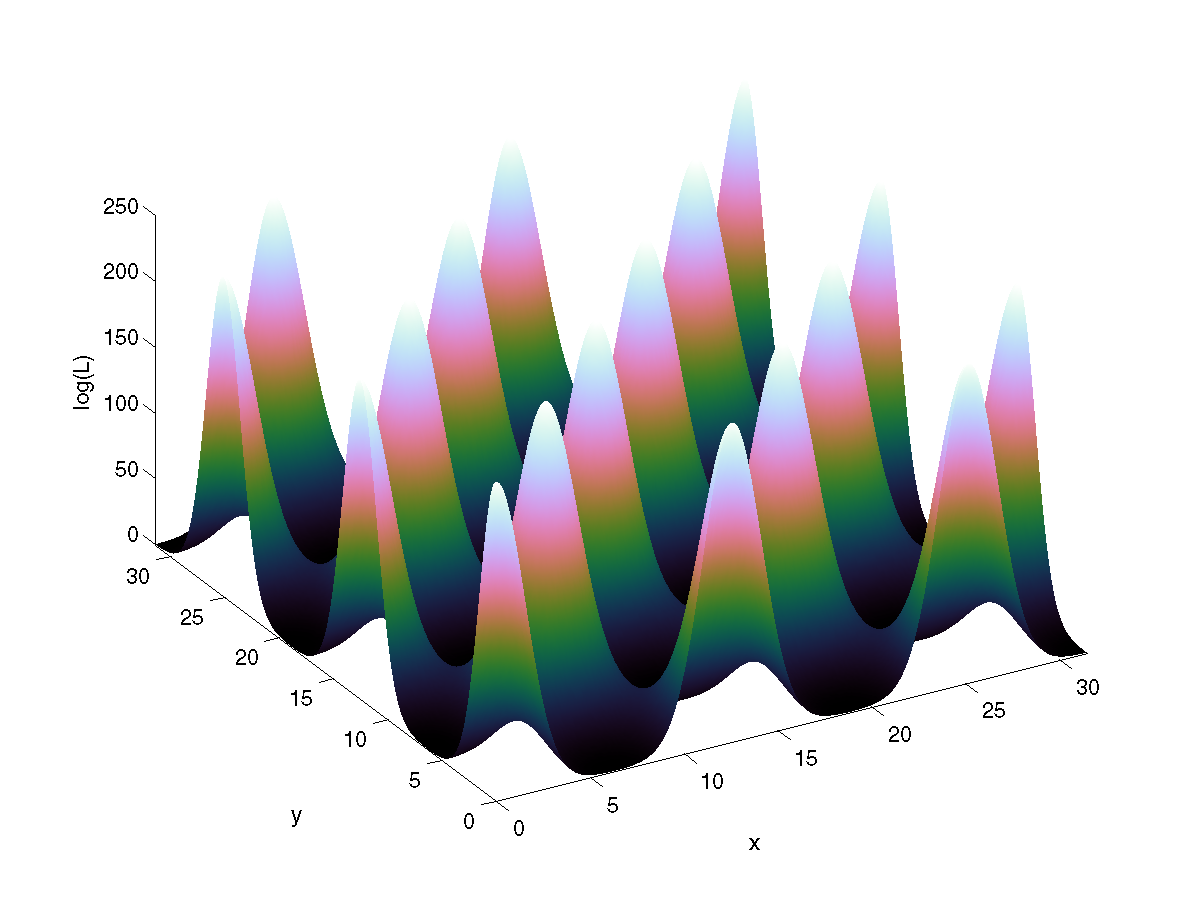}
\caption{The eggbox log-likelihood surface, given by Equation~\eqref{eq:eggboxlike}.}
\label{fig:eggboxlike}
\end{center}
\end{figure}

This is a standard example of a very multimodal likelihood distribution in two dimensions. It has many peaks of equal value, so the network must be able to take samples from separated regions of the prior and make accurate predictions in all peaks. The eggbox likelihood~\citep{MultiNest2} is given by
\begin{equation}
\label{eq:eggboxlike}
\mathcal{L}(x,y) = \exp \left[ \left( 2 + \cos(\tfrac{x}{2})\cos(\tfrac{y}{2}) \right)^5 \right],
\end{equation}
where we take a uniform prior $\mathcal{U}(0,10\pi)$ for both $x$ and $y$. The structure of the surface can be seen in Figure~\ref{fig:eggboxlike}.

\begin{figure}
\begin{center}
\subfigure[]{\includegraphics[width=0.45\columnwidth]{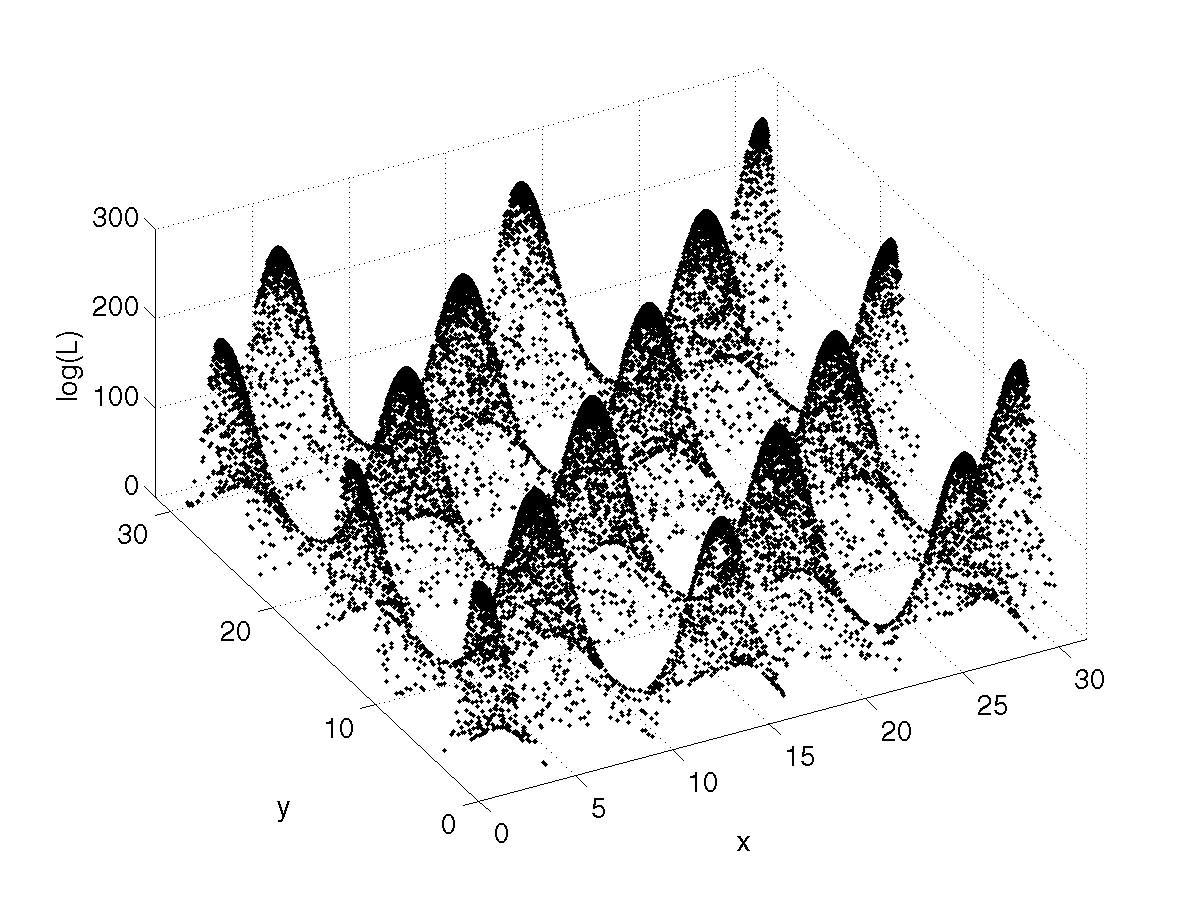}}\hspace{0.5cm}
\subfigure[]{\includegraphics[width=0.45\columnwidth]{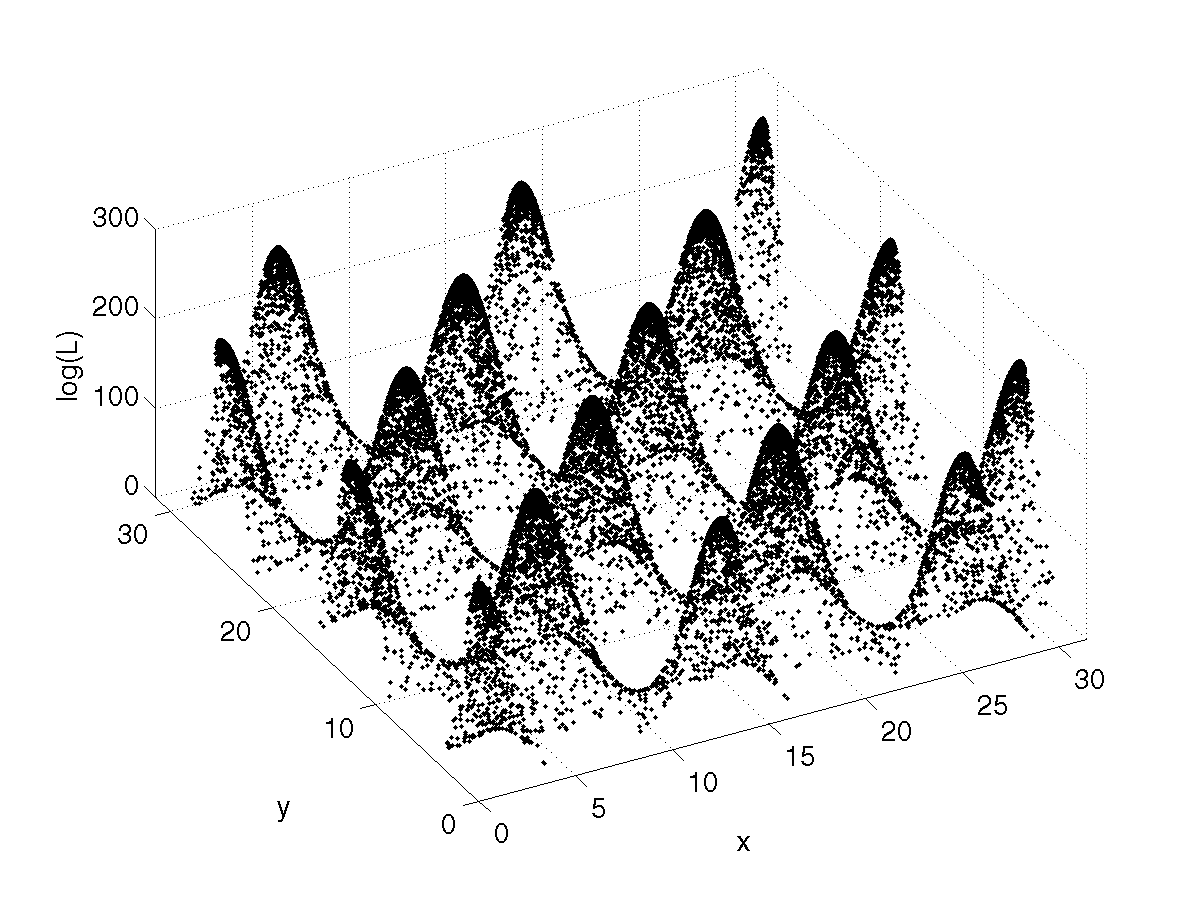}}
\caption{Points of lowest likelihood of the eggbox likelihood from successive iterations as given by (a) {\sc MultiNest} and (b) BAMBI.}
\label{fig:eggboxpost}
\end{center}
\end{figure}

We ran the eggbox example in both {\sc MultiNest} and BAMBI, both using $4000$ live points. For BAMBI, we used $4000$ samples for training a network with $50$ hidden nodes. In Table~\ref{tab:eggboxev} we report the evidences recovered by both methods as well as the true value obtained analytically from Equation~\eqref{eq:eggboxlike}. Both methods return evidences that agree with the analytically determined value to within the given error bounds. Figure~\ref{fig:eggboxpost} compares the posterior probability distributions returned by the two algorithms via the distribution of lowest-likelihood points removed at successive iterations by {\sc MultiNest}. We can see that they are identical distributions; therefore, we can say that the use of the NN did not reduce the quality of the results either for parameter estimation or model selection. During the BAMBI analysis $51.3\%$ of the log-likelihood function evaluations were done using the NN; if this were a more computationally expensive function, significant speed gains would have been realised.
\begin{table}
\begin{center}
\begin{tabular}{c|cc}
Method & $\log(\mathcal{Z})$ \\ \hline \hline
Analytical & $235.88$ \\
{\sc MultiNest} & $235.859 \pm 0.039$ \\
BAMBI & $235.901 \pm 0.039$ \\
\end{tabular}
\end{center}
\caption{The log-evidence values of the eggbox likelihood as found analytically and with {\sc MultiNest} and BAMBI.}
\label{tab:eggboxev}
\end{table}

\subsection{Gaussian Shells}\label{sec:gaussianshell}
\begin{figure}
\begin{center}
\includegraphics[width=3in]{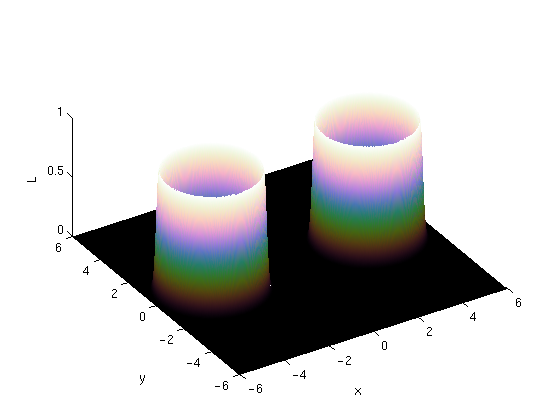}
\caption{The Gaussian shell likelihood surface, given by Equations~\eqref{eq:gausslike} and~\eqref{eq:circfctn}.}
\label{fig:gausslike}
\end{center}
\end{figure}

The Gaussian shells likelihood function has low values over most of the prior, except for thin circular shells that have Gaussian cross-sections. We use two separate Gaussian shells of equal magnitude so that this is also a mutlimodal inference problem. Therefore, our Gaussian shells likelihood is
\begin{equation}
\label{eq:gausslike}
\mathcal{L}({\bf x}) = \textrm{circ}({\bf x};{\bf c}_1,r_1,w_1)+\textrm{circ}({\bf x};{\bf c}_2,r_2,w_2),
\end{equation}
where each shell is defined by
\begin{equation}
\label{eq:circfctn}
\textrm{circ}({\bf x};{\bf c},r,w) = \frac{1}{\sqrt{2\pi w^2}} \exp\left[-\frac{(\lvert{\bf x}-{\bf c}\rvert-r)^2}{2w^2}\right].
\end{equation}
This is shown in Figure~\ref{fig:gausslike}.

\begin{figure}
\begin{center}
\subfigure[]{\includegraphics[width=0.45\columnwidth]{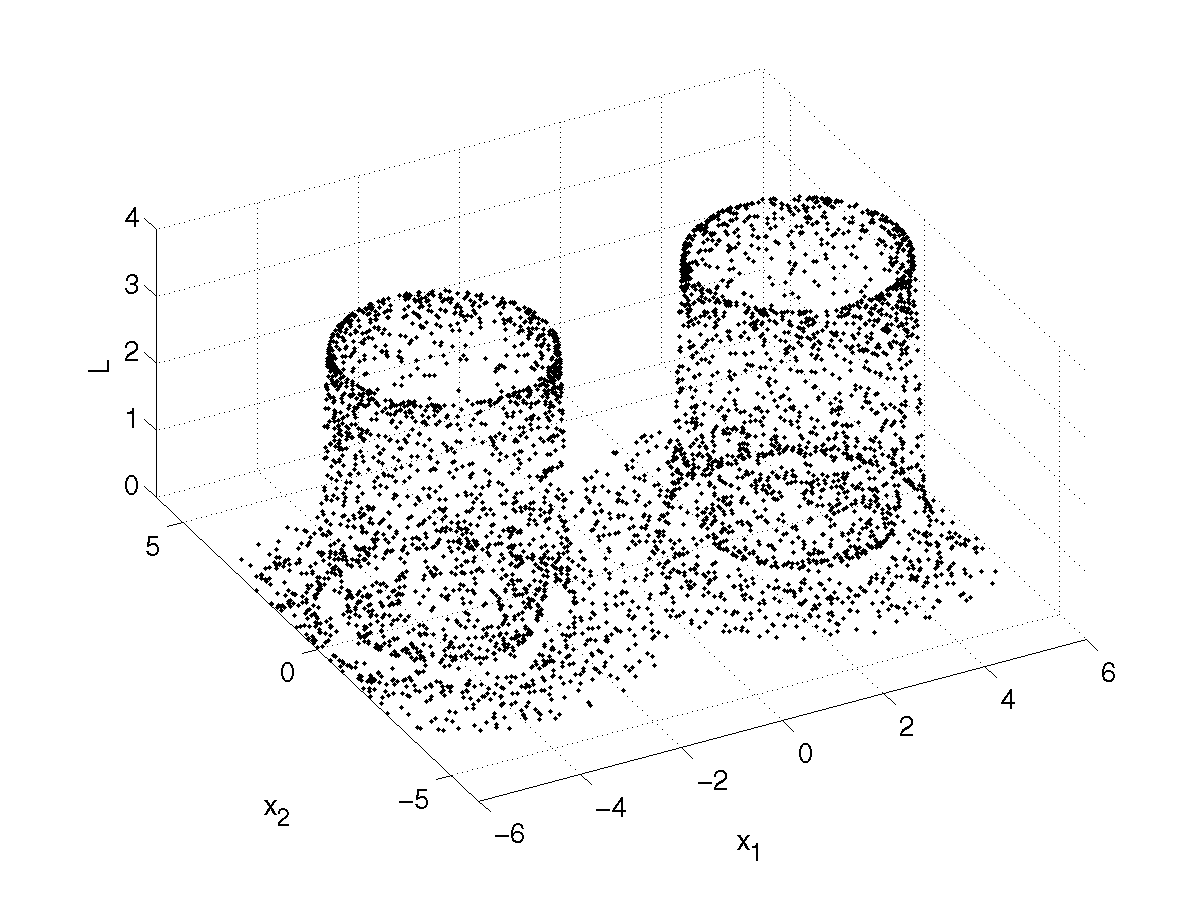}}\hspace{0.5cm}
\subfigure[]{\includegraphics[width=0.45\columnwidth]{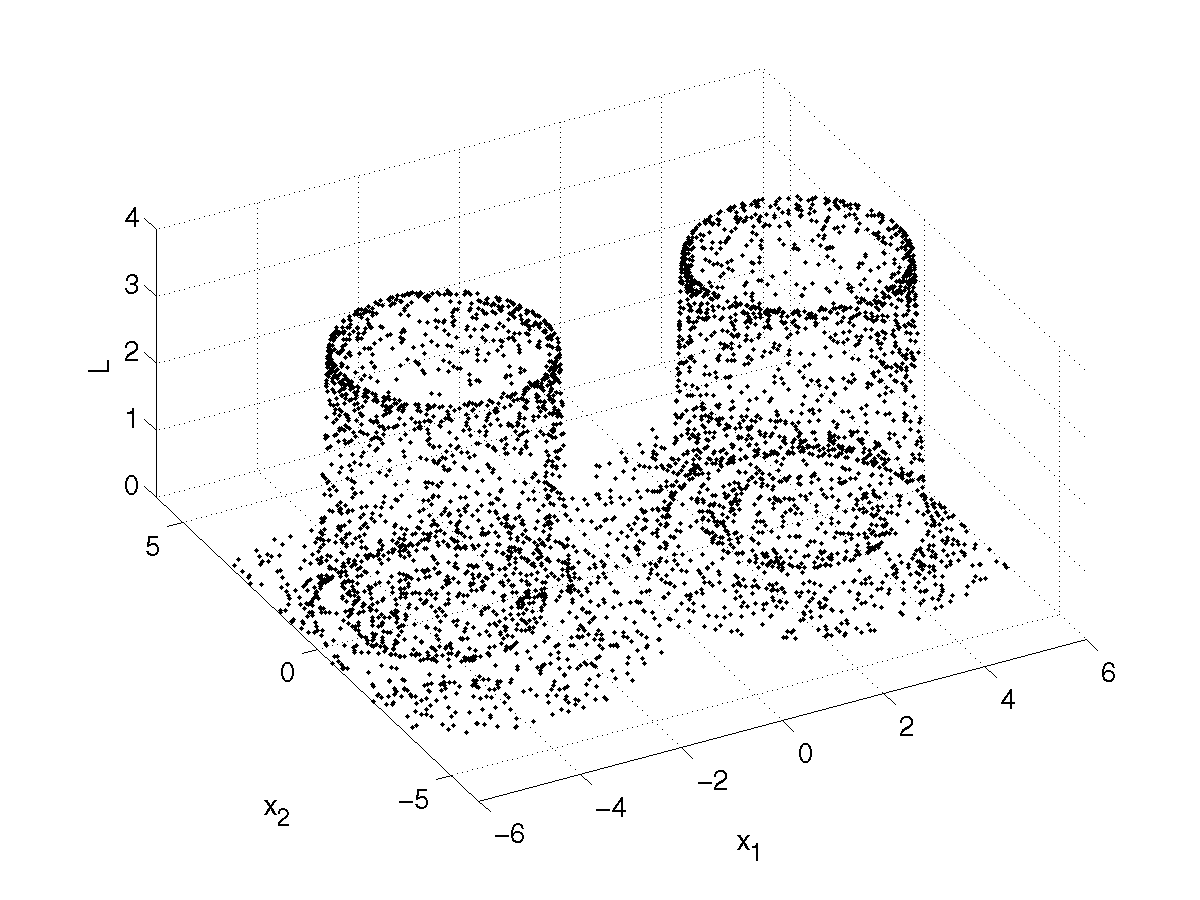}}
\caption{Points of lowest likelihood of the Gaussian shell likelihood from successive iterations as given by (a) {\sc MultiNest} and (b) BAMBI.}
\label{fig:gausspost}
\end{center}
\end{figure}

As with the eggbox problem, we analysed the Gaussian shells likelihood with both {\sc MultiNest} and BAMBI using uniform priors $\mathcal{U}(-6,6)$ on both dimensions of ${\bf x}$. {\sc MultiNest} sampled with $1000$ live points and BAMBI used $2000$ samples for training a network with $100$ hidden nodes. In Table~\ref{tab:gaussev} we report the evidences recovered by both methods as well as the true value obtained analytically from Equations~\eqref{eq:gausslike} and~\eqref{eq:circfctn}. The evidences are both consistent with the true value. Figure~\ref{fig:gausspost} compares the posterior probability distributions returned by the two algorithms (in the same manner as with the eggbox example). Again, we see that the distribution of returned values is nearly identical when using the NN, which BAMBI used for $18.2\%$ of its log-likelihood function evaluations. This is a significant fraction, especially since they are all at the end of the analysis when exploring the peaks of the distribution.
\begin{table}
\begin{center}
\begin{tabular}{c|cc}
Method & $\log(\mathcal{Z})$ \\ \hline \hline
Analytical & $-1.75$ \\
{\sc MultiNest} & $-1.768 \pm 0.052$ \\
BAMBI & $-1.757 \pm 0.052$ \\ 
\end{tabular}
\end{center}
\caption{The log-evidence values of the Gaussian shell likelihood as found analytically and with {\sc MultiNest} and BAMBI.}
\label{tab:gaussev}
\end{table}

\subsection{Rosenbrock function}\label{sec:rosenbrock}
\begin{figure}
\begin{center}
\includegraphics[width=3in]{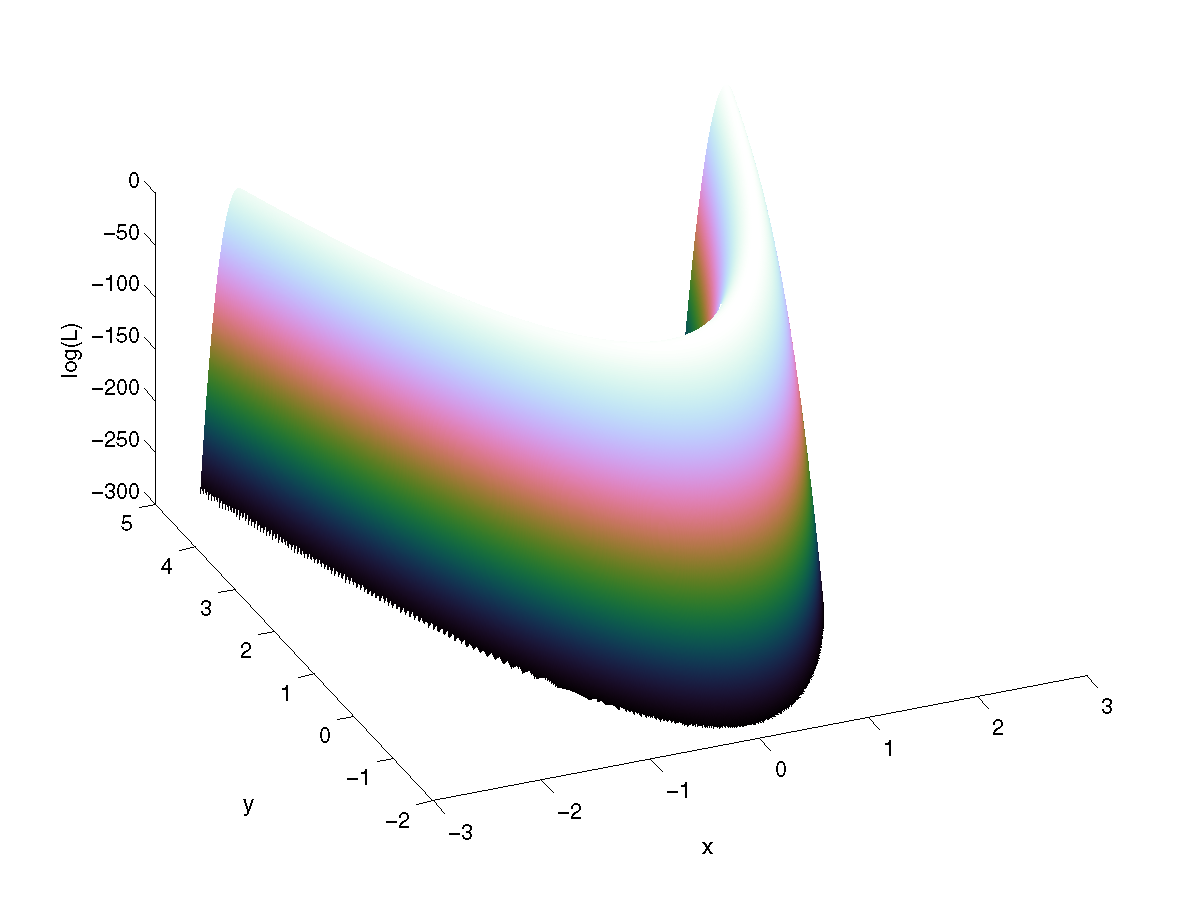}
\caption{The Rosenbrock log-likelihood surface given by Equation~\eqref{eq:rosenbrocklike} with $N=2$.}
\label{fig:rosenbrocklike}
\end{center}
\end{figure}

The Rosenbrock function is a standard example used for testing optimization as it presents a long, curved degeneracy through all dimensions. For our NN training, it presents the difficulty of learning the likelihood function over a large, curving region of the prior. We use the Rosenbrock function to define the negative log-likelihood, so the likelihood function is given in $N$ dimensions by
\begin{equation}
\mathcal{L}({\bf x}) = \exp \left\{-\sum_{i=1}^{N-1}{\left[(1-x_i)^2 + 100 (x_{i+1}-x_i^2)^2\right]}\right\}.
\label{eq:rosenbrocklike}
\end{equation}
Figure~\ref{fig:rosenbrocklike} shows how this appears for $N=2$.

\begin{figure}
\begin{center}
\subfigure[]{\includegraphics[width=0.45\columnwidth]{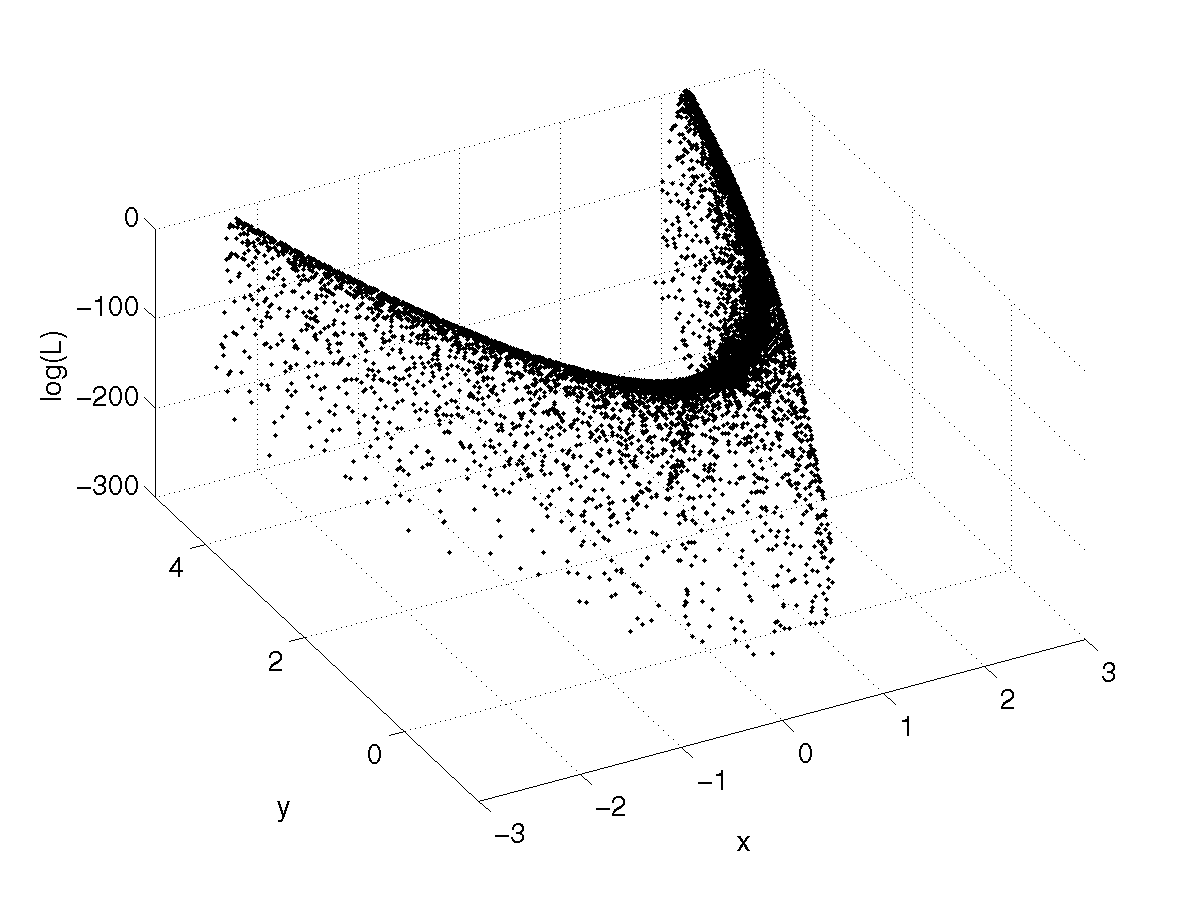}}\hspace{0.5cm}
\subfigure[]{\includegraphics[width=0.45\columnwidth]{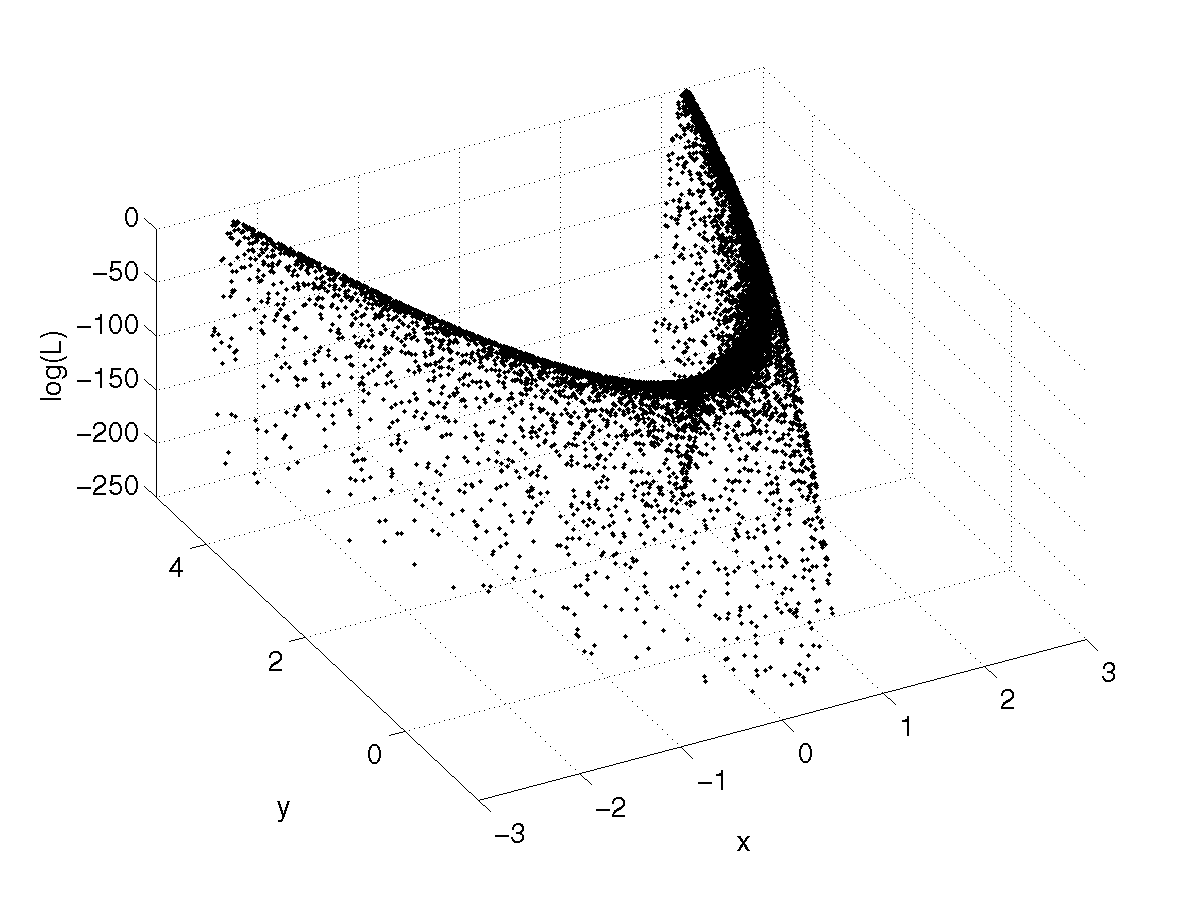}}
\caption{Points of lowest likelihood of the Rosenbrock likelihood for $N=2$ from successive iterations as given by (a) {\sc MultiNest} and (b) BAMBI.}
\label{fig:rosenbrockpost}
\end{center}
\end{figure}

\begin{figure}
\begin{center}
\includegraphics[width=3in]{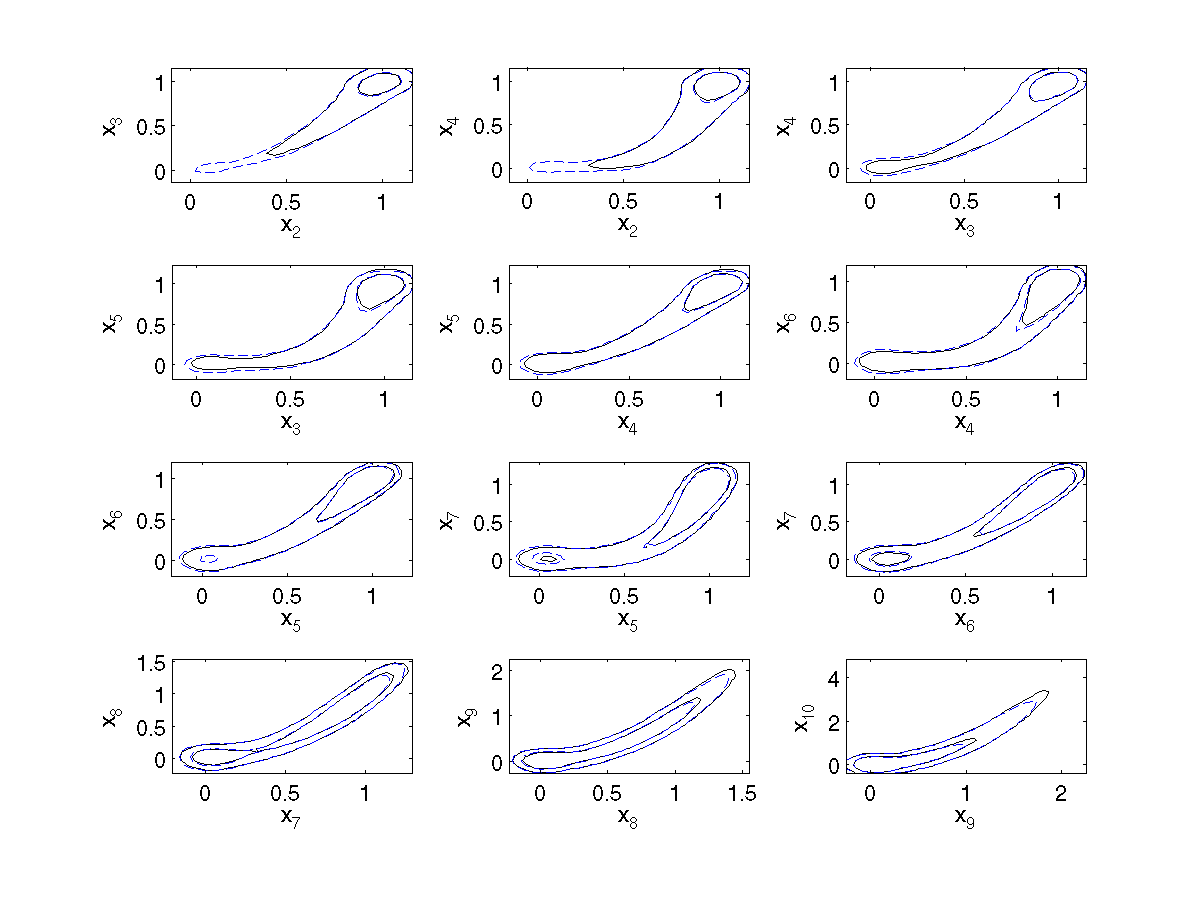}
\caption{Marginalised 2D posteriors for the Rosenbrock function with $N=10$. The $12$ most correlated pairs are shown. {\sc MultiNest} is in solid black, BAMBI in dashed blue. Inner and outer contours represent $68\%$ and $95\%$ confidence levels, respectively.}
\label{fig:rosenbrock10post}
\end{center}
\end{figure}

We set uniform priors of $\mathcal{U}(-5,5)$ in all dimensions and performed analysis with both {\sc MultiNest} and BAMBI with $N=2$ and $N=10$. For $N=2$, {\sc MultiNest} sampled with $2000$ live points and BAMBI used $2000$ samples for training a NN with $50$ hidden-layer nodes. With $N=10$, we used $2000$ live points, $6000$ samples for network training, and $50$ hidden nodes. Table~\ref{tab:rosenbrockev} gives the calculated evidences values returned by both algorithms as well as the analytically calculated values from Equation~\eqref{eq:rosenbrocklike} (there does not exist an analytical solution for the 10D case so this is not included). Figure~\ref{fig:rosenbrockpost} compares the posterior probability distributions returned by the two algorithms for $N=2$. For $N=10$, we show in Figure~\ref{fig:rosenbrock10post} comparisons of the marginalised two-dimensional posterior distributions for $12$ variable pairs. We see that {\sc MultiNest} and BAMBI return nearly identical posterior distributions as well as consistent estimates of the evidence. For $N=2$ and $N=10$, BAMBI was able to use a NN for $64.7\%$ and $30.5\%$ of its log-likelihood evaluations respectively. Even when factoring in time required to train the NN this would have resulted in large decreases in running time for a more computationally expensive likelihood function.
\begin{table}
\begin{center}
\begin{tabular}{c|cc}
Method & $\log(\mathcal{Z})$ \\ \hline \hline
Analytical 2D & $-5.804$ \\
{\sc MultiNest} 2D & $-5.799 \pm 0.049$ \\
BAMBI 2D & $-5.757 \pm 0.049$ \\ \hline
{\sc MultiNest} 10D & $-41.54 \pm 0.13$ \\
BAMBI 10D & $-41.53 \pm 0.13$ \\
\end{tabular}
\end{center}
\caption{The log-evidence values of the Rosenbrock likelihood as found analytically and with {\sc MultiNest} and BAMBI.}
\label{tab:rosenbrockev}
\end{table}

\section{Cosmological Parameter Estimation with BAMBI}\label{sec:cosmology}
While likelihood functions resembling our previous toy examples do exist in real physical models, we would also like to demonstrate the usefulness of BAMBI on simpler likelihood surfaces where the time of evaluation is the critical limiting factor. One such example in astrophysics is that of cosmological parameter estimation and model selection.

We implement BAMBI within the standard {\sc CosmoMC} code~\citep{CosmoMC}, which by default uses MCMC sampling. This allows us to compare the performance of BAMBI with other methods, such as {\sc MultiNest}~\citep{MultiNest2}, {\sc CosmoNet}~\citep{CosmoNet1,CosmoNet2}, {\sc InterpMC}~\citep{InterpMC}, PICO~\citep{PICO}, and others. In this paper, we will only report performances of BAMBI and {\sc MultiNest}, but these can be compared with reported performance from the other methods.

Bayesian parameter estimation in cosmology requires evaluation of theoretical temperature and polarisation CMB power spectra ($C_l$ values) using codes such as CAMB~\citep{CAMB}. These evaluations can take on the order of tens of seconds depending on the cosmological model. The $C_l$ spectra are then compared to WMAP and other observations for the likelihood function. Considering that thousands of these evaluations will be required, this is a computationally expensive step and a limiting factor in the speed of any Bayesian analysis. With BAMBI, we use samples to train a NN on the combined likelihood function which will allow us to forgo evaluating the full power spectra. This has the benefit of not requiring a pre-computed sample of points as in {\sc CosmoNet} and PICO, which is particularly important when including new parameters or new physics in the model. In these cases we will not know in advance where the peak of the likelihood will be and it is around this location that the most samples would be needed for accurate results.

The set of cosmological parameters that we use as variables and their prior ranges is given in Table~\ref{tab:cosmoparams}; the parameters have their usual meanings in cosmology~\citep[see][table 1]{WMAP7}. The prior probability distributions are uniform over the ranges given. A non-flat cosmological model incorporates all of these parameters, while we set $\Omega_K=0$ for a flat model. We use $w=-1$ in both cases. The flat model thus represents the standard $\Lambda$CDM cosmology. We use two different data sets for analysis:  ($1$) CMB observations alone and ($2$) CMB observations plus Hubble Space Telescope constraints on $H_0$, large-scale structure constraints from the luminous red galaxy subset of the SDSS and the 2dF survey, and supernovae Ia data. The CMB dataset consists of WMAP seven-year data~\citep{WMAP7} and higher resolution observations from the ACBAR, CBI, and BOOMERanG experiments. The {\sc CosmoMC} website~\citep[see][]{CosmoMC} provides full references for the most recent sources of these data.
\begin{table}
\begin{center}
\begin{tabular}{ccc}
Parameter & Min & Max \\ \hline \hline
$\Omega_b h^2$ & $0.018$ & $0.032$ \\
$\Omega_{DM} h^2$ & $0.04$ & $0.16$ \\
$\theta$ & $0.98$ & $1.1$ \\
$\tau$ & $0.01$ & $0.5$ \\
$\Omega_K$ & $-0.1$ & $0.1$ \\
$n_s$ & $0.8$ & $1.2$ \\
$\log[10^{10} A_s]$ & $2.7$ & $4$ \\
$A_{SZ}$ & $0$ & $2$ \\
\end{tabular}
\caption{The cosmological parameters and their minimum and maximum values. Uniform priors were used on all variables. $\Omega_K$ was set to $0$ for the flat model.}
\label{tab:cosmoparams}
\end{center}
\end{table}

\begin{figure}
\begin{center}
\includegraphics[width=3in]{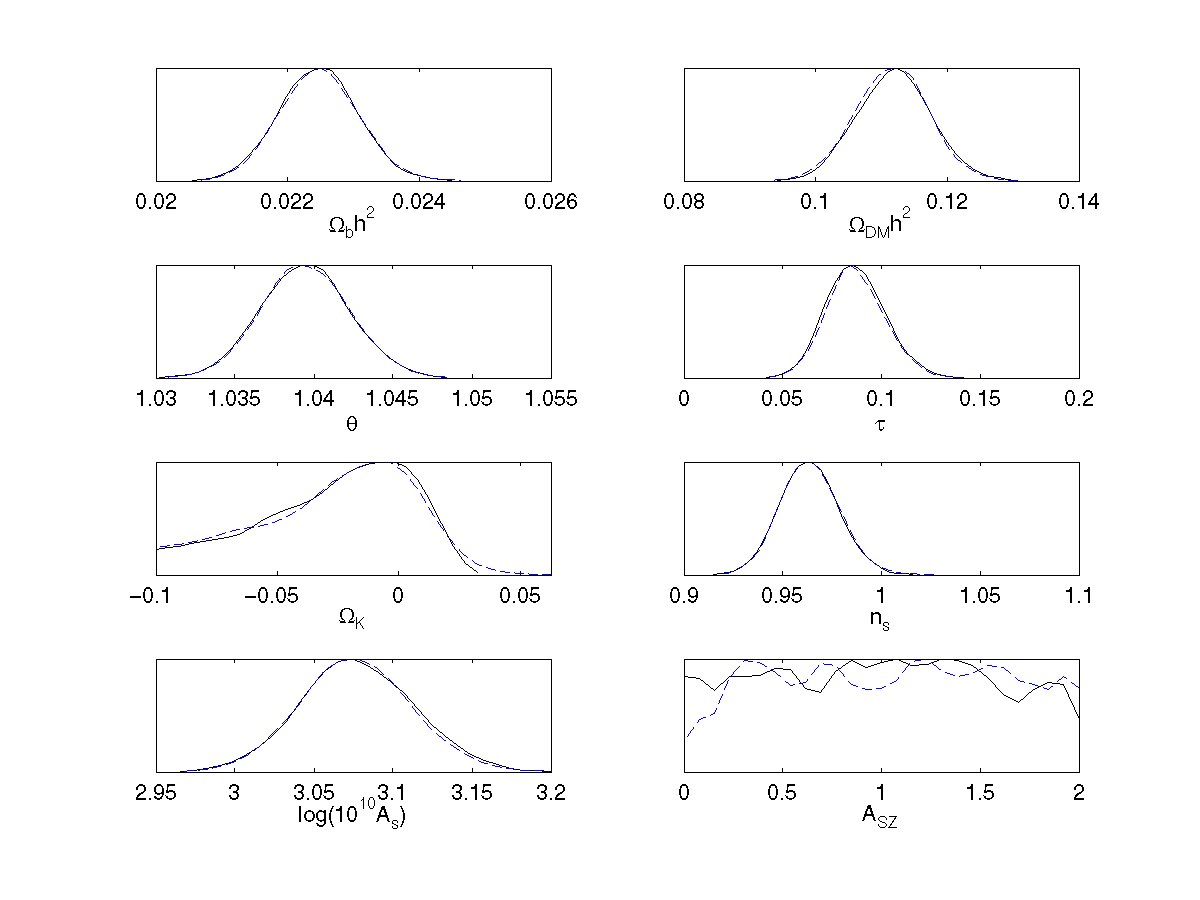}
\caption{Marginalised 1D posteriors for the non-flat model ($\Lambda$CDM$+\Omega_K$) using only CMB data. {\sc MultiNest} is in solid black, BAMBI in dashed blue.}
\label{fig:nonflat_cmb_1d}
\end{center}
\end{figure}

\begin{figure}
\begin{center}
\includegraphics[width=3in]{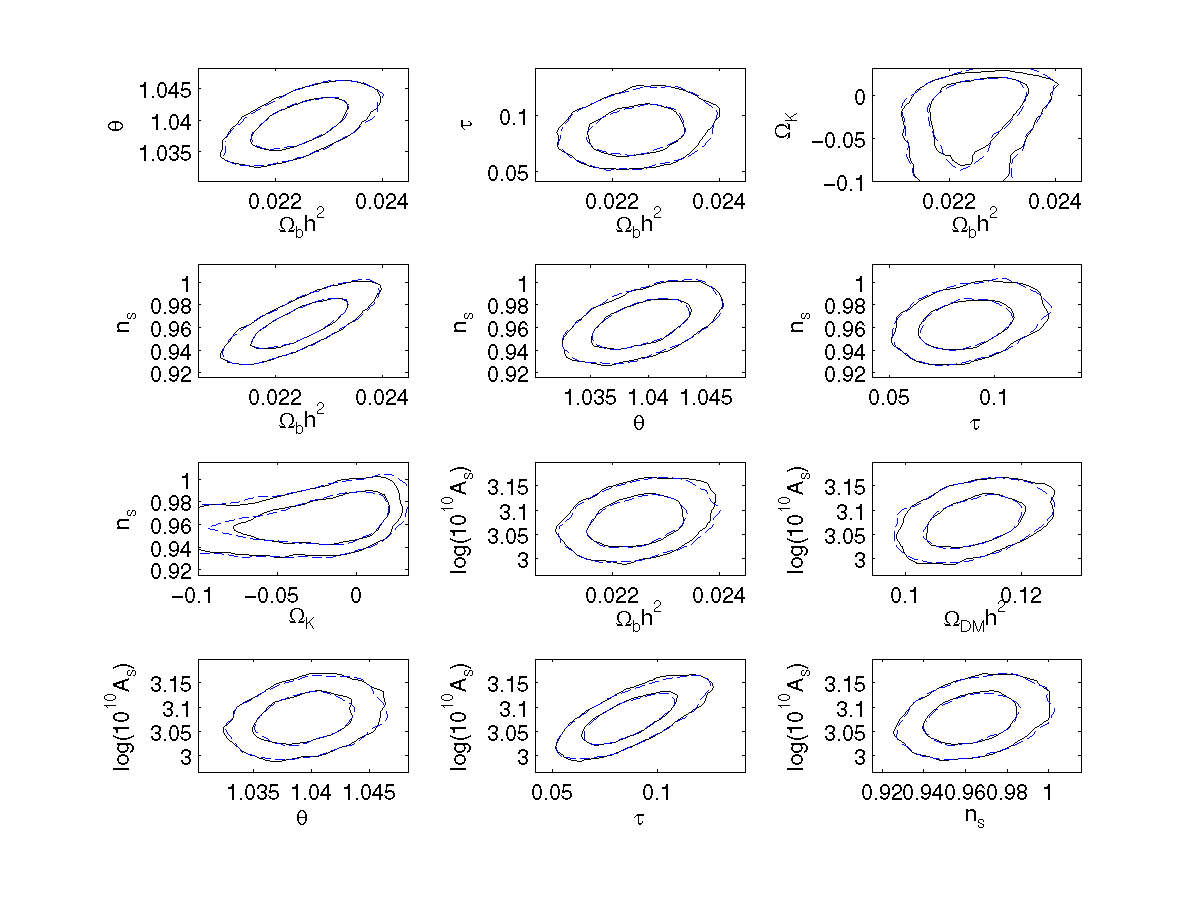}
\caption{Marginalised 2D posteriors for the non-flat model ($\Lambda$CDM$+\Omega_K$) using only CMB data. The $12$ most correlated pairs are shown. {\sc MultiNest} is in solid black, BAMBI in dashed blue. Inner and outer contours represent $68\%$ and $95\%$ confidence levels, respectively.}
\label{fig:nonflat_cmb_2d}
\end{center}
\end{figure}

\begin{figure}
\begin{center}
\includegraphics[width=3in]{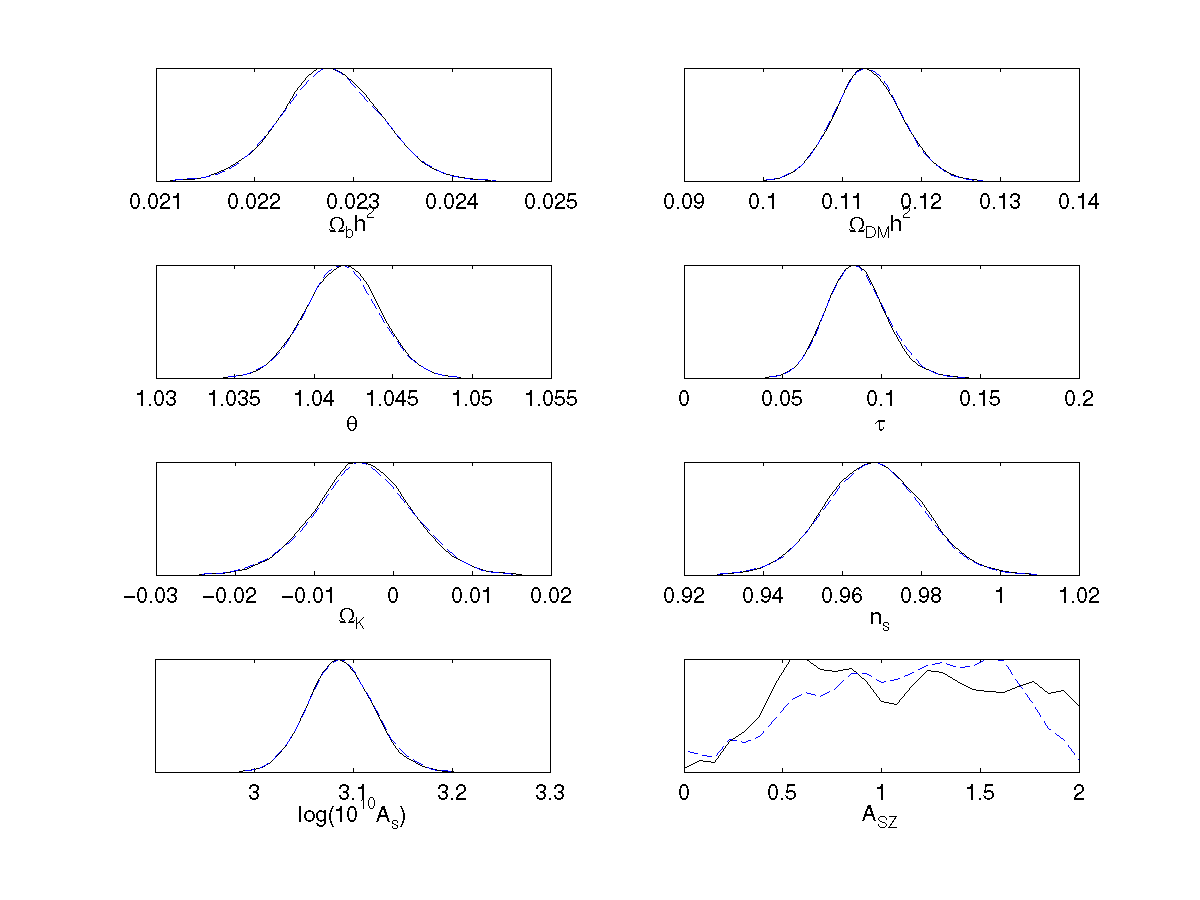}
\caption{Marginalised 1D posteriors for the non-flat model ($\Lambda$CDM$+\Omega_K$) using the complete data set. {\sc MultiNest} is in solid black, BAMBI in dashed blue.}
\label{fig:nonflat_all_1d}
\end{center}
\end{figure}

\begin{figure}
\begin{center}
\includegraphics[width=3in]{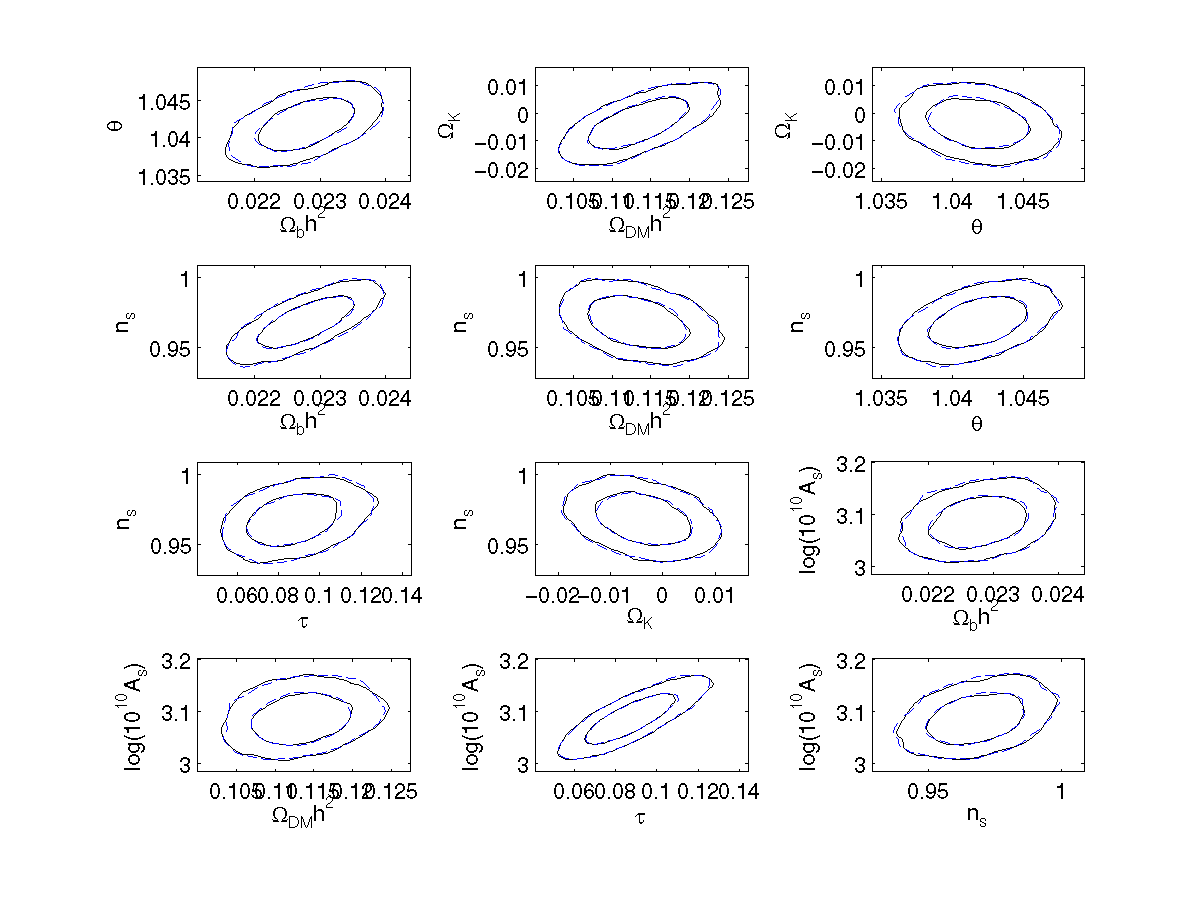}
\caption{Marginalised 2D posteriors for the non-flat model ($\Lambda$CDM$+\Omega_K$) using the complete data set. The $12$ most correlated pairs are shown. {\sc MultiNest} is in solid black, BAMBI in dashed blue. Inner and outer contours represent $68\%$ and $95\%$ confidence levels, respectively.}
\label{fig:nonflat_all_2d}
\end{center}
\end{figure}

Analyses with {\sc MultiNest} and BAMBI were run on all four combinations of models and data sets. {\sc MultiNest} sampled with $1000$ live points and an efficiency of $0.5$, both on its own and within BAMBI; BAMBI used $2000$ samples for training a NN on the likelihood, with $50$ hidden-layer nodes for both the flat model and non-flat model. Table~\ref{tab:cosmo_evidences} reports the recovered evidences from the two algorithms for both models and both data sets. It can be seen that the two algorithms report equivalent values to within statistical error for all four combinations. In Figures~\ref{fig:nonflat_cmb_1d} and~\ref{fig:nonflat_cmb_2d} we show the recovered one- and two-dimensional marginalised posterior probability distributions for the non-flat model using the CMB-only data set. Figures~\ref{fig:nonflat_all_1d} and~\ref{fig:nonflat_all_2d} show the same for the non-flat model using the complete data set. We see very close agreement between {\sc MultiNest} (in solid black) and BAMBI (in dashed blue) across all parameters. The only exception is $A_{SZ}$ since it is unconstrained by these models and data and is thus subject to large amounts of variation in sampling. The posterior probability distributions for the flat model with either data set are extremely similar to those of the non-flat flodel with setting $\Omega_K=0$, as expected, so we do not show them here.
\begin{table}
\begin{center}
\begin{tabular}{cccc}
Algorithm & Model & Data Set & $\log(\mathcal{Z})$ \\ \hline \hline
{\sc MultiNest} & $\Lambda$CDM & CMB only & $-3754.58 \pm 0.12$ \\
BAMBI & $\Lambda$CDM & CMB only & $-3754.57 \pm 0.12$ \\ \hline
{\sc MultiNest} & $\Lambda$CDM & all & $-4124.40 \pm 0.12$ \\
BAMBI & $\Lambda$CDM & all & $-4124.11 \pm 0.12$ \\ \hline
{\sc MultiNest} & $\Lambda$CDM$+\Omega_K$ & CMB only & $-3755.26 \pm 0.12$ \\
BAMBI & $\Lambda$CDM$+\Omega_K$ & CMB only & $-3755.57 \pm 0.12$ \\ \hline
{\sc MultiNest} & $\Lambda$CDM$+\Omega_K$ & all & $-4126.54 \pm 0.13$ \\
BAMBI & $\Lambda$CDM$+\Omega_K$ & all & $-4126.35 \pm 0.13$ \\
\end{tabular}
\caption{Evidences calculated by {\sc Multinest} and BAMBI for the flat ($\Lambda$CDM) and non-flat ($\Lambda$CDM$+\Omega_K$) models using the CMB-only and complete data sets. The two algorithms are in close agreement in all cases.}
\label{tab:cosmo_evidences}
\end{center}
\end{table}


\begin{figure}
\begin{center}
\includegraphics[width=3in]{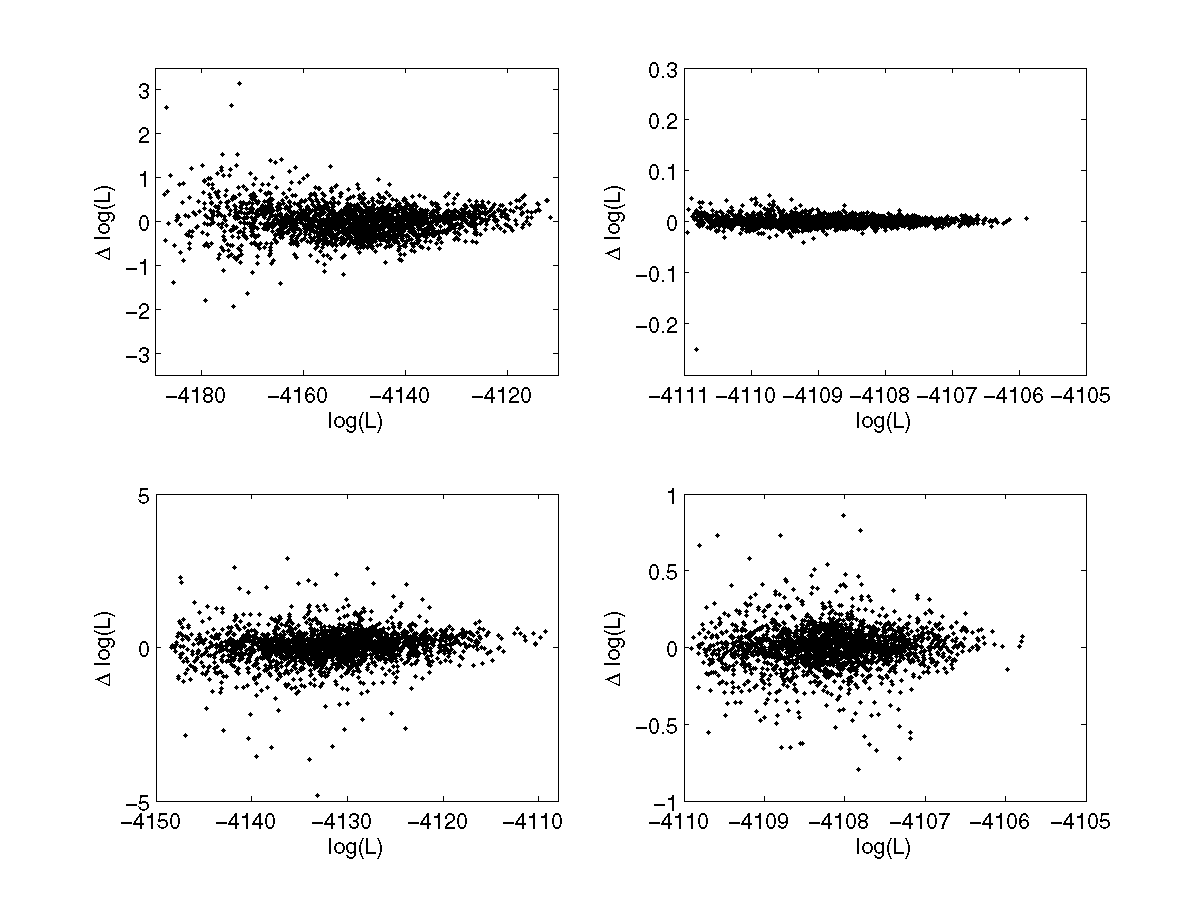}
\caption{The error in the predicted likelihood ($\Delta \log(\mathcal{L}) = \log(\mathcal{L}_{\textrm{predicted}}) - \log(\mathcal{L}_{\textrm{true}})$) for the BAMBI networks trained on the flat (top row) and non-flat (bottom row) models using the complete data set. The left column represents predictions from the first NN trained to sufficient accuracy; the right column are results from the second, and final, NN trained in each case. The flat and non-flat models both used $50$ hidden-layer nodes.}
\label{fig:networkaccuracy}
\end{center}
\end{figure}

A by-product of running BAMBI is that we now have a network that is trained to predict likelihood values near the peak of the distribution. To see how accurate this network is, in Figure~\ref{fig:networkaccuracy} we plot the error in the prediction ($\Delta \log(\mathcal{L}) = \log(\mathcal{L}_{\textrm{predicted}}) - \log(\mathcal{L}_{\textrm{true}})$) versus the true log-likelihood value for the different sets of training and validation points that were used. What we show are results for networks that were trained to sufficient accuracy in order to be used for making likelihood predictions; this results in two networks for each case shown. We can see that although the flat model used the same number of hidden-layer nodes, the simpler physical model allowed for smaller error in the likelihood predictions. Both final networks (one for each model) are significantly more accurate than the specified tolerance of a maximum standard deviation on the error of $0.5$. In fact, for the flat model, all but one of the $2000$ points have an error of less than $0.06$ log-units. The two networks trained in each case overlap in the range of log-likelihood values on which they trained. The first network, although trained to lower accuracy, is valid over a much larger range of log-likelihoods. The accuracy of each network increases with increasing true log-likelihood and the second network, trained on higher log-likelihood values, is significantly more accurate than the first.

The final comparison, and perhaps the most important, is the running time required. The analyses were run using MPI parallelisation on $48$ processors. We recorded the time required for the complete analysis, not including any data initialisation prior to initial sampling. We then divide this time by the number of likelihood evaluations performed to obtain an average time per likelihood ($t_{\textrm{wall clock, sec}} \times N_{\textrm{CPUs}} / N_{\log(\mathcal{L}) \textrm{ evals}}$). Therefore, time required to train the NN is still counted as a penalty factor. If a NN takes more time to train, this will hurt the average time, but obtaining a usable NN sooner and with fewer training calls will give a better time since more likelihoods will be evaluated by the NN. The resulting average times per likelihood and speed increases are given in Table~\ref{tab:cosmo_speed}. Although the speed increases appear modest, one must remember that these include time taken to train the NNs, during which no likelihoods were evaluated. This can be seen in that although $30$--$40\%$ of likelihoods are evaluated with a NN, as reported in Table~\ref{tab:cosmo_speed2}, we do not obtain the full equivalent speed increase. We are still able to obtain a significant decrease in running time while adding in the bonus of having a NN trained on the likelihood function.
\begin{table}
\begin{center}
\begin{tabular}{ccccc}
Model & Data set & {\sc MultiNest} & BAMBI & Speed \\ 
 & & $t_{\mathcal{L}}$ (s) & $t_{\mathcal{L}}$ (s) & factor \\ \hline \hline
$\Lambda$CDM & CMB only & $2.394$ & $1.902$ & $1.26$ \\
$\Lambda$CDM & all & $3.323$ & $2.472$ & $1.34$ \\
$\Lambda$CDM$+\Omega_K$ & CMB only & $12.744$ & $9.006$ & $1.42$ \\
$\Lambda$CDM$+\Omega_K$ & all & $12.629$ & $10.651$ & $1.19$ \\
\end{tabular}
\caption{Time per likelihood evaluation, factor of speed increase from {\sc MultiNest} to BAMBI ($t_{\textrm{MN}}/t_{\textrm{BAMBI}}$).}
\label{tab:cosmo_speed}
\end{center}
\end{table}
\begin{table}
\begin{center}
\begin{tabular}{ccccc}
Model & Data set & $\% \log(\mathcal{L})$ & Equivalent & Actual \\ 
 & & with NN & speed factor & speed factor \\ \hline \hline
$\Lambda$CDM & CMB only & $40.5$ & $1.68$ & $1.26$ \\
$\Lambda$CDM & all & $40.2$ & $1.67$ & $1.34$ \\
$\Lambda$CDM$+\Omega_K$ & CMB only & $34.2$ & $1.52$ & $1.42$ \\
$\Lambda$CDM$+\Omega_K$ & all & $30.0$ & $1.43$ & $1.19$ \\
\end{tabular}
\caption{Percentage of likelihood evaluations performed with a NN, equivalent speed factor, and actual factor of speed increase.}
\label{tab:cosmo_speed2}
\end{center}
\end{table}

\section{Using Trained Networks for Follow-up in BAMBI}\label{sec:NNanalysis}
A major benefit of BAMBI is that following an initial run the user is provided with a trained NN, or multiple ones, that model the log-likelihood function. These can be used in a subsequent analysis with different priors to obtain much faster results. This is a comparable analysis to that of {\sc CosmoNet}~\citep{CosmoNet1,CosmoNet2}, except that the NNs here are a product of an initial Bayesian analysis where the peak of the distribution was {\em not} previously known. No prior knowledge of the structure of the likelihood surface was used to generate the networks that are now able to be re-used.

When multiple NNs are trained and used in the initial BAMBI analysis, we must determine which network's prediction to use in the follow-up. The approximate error of uncertainty of a NN's prediction of the value $y({\bf x}; {\bf a})$ (${\bf x}$ denoting input parameters, ${\bf a}$ NN weights and biases as before) that models the log-likelihood function is given by~\cite{NNprederror} as
\begin{equation}
\sigma^2 = \sigma_{\textrm{pred}}^2 + \sigma_{\nu}^2,
\label{eq:NNerrorbar1}
\end{equation}
where
\begin{equation}
\sigma_{\textrm{pred}}^2 = {\bf g}^{\textrm{T}} \bsf{B}^{-1} {\bf g}
\label{eq:NNerrorbar2}
\end{equation}
and $\sigma_{\nu}^2$ is the variance of the noise on the output from the network training. In Equation~\eqref{eq:NNerrorbar2}, $\bsf{B}$ is the Hessian of the log-posterior as before, and ${\bf g}$ is the gradient of the NN's prediction with respect to the weights about their maximum posterior values,
\begin{equation}
{\bf g} = \frac{\partial y({\bf x}; {\bf a})}{\partial {\bf a}} \rvert_{{\bf x},{\bf a}_{\textrm{MP}}}.
\label{eq:predgrad}
\end{equation}
This uncertainty of the prediction is calculated for each training and validation data point used in the initial training of the NN for each saved NN. The threshold for accepting a predicted point from a network is then set to be $1.2$ times the maximum uncertainty found.

A log-likelihood is calculated by first making a prediction with the final NN to be trained and saved and then calculating the error for this prediction. If the error, $\sigma_{\textrm{pred}}$, is greater than that NN's threshold, then we consider the previous trained NN. We again calculate the predicted log-likelihood value and error to compare with its threshold. This continues to the first NN saved until a NN makes a sufficiently confident prediction. If no NNs can make a confident enough prediction, then we set $\log(\mathcal{L})=-\infty$. This is justified because the NNs are trained to predict values in the highest likelihood regions of parameter space and if a set of parameters lies outside their collective region of validity, then it must not be within the region of highest likelihoods.

To demonstrate the speed-up potential of using the NNs, we first ran an analysis of the cosmological parameter estimation using both models and both data sets. This time, however, we set the tolerance of the NNs to $1.0$ instead of $0.5$, so that they would be valid over a larger range of log-likelihoods and pass the accuracy criterion sooner. Each analysis produced two trained NNs. We then repeated each of the four analyses, but set the prior ranges to be uniform over the region defined by ${\bf x}_{\textrm{max}(\log(\mathcal{L}))} \pm 4\bsigma$, where $\bsigma$ is the vector of standard deviations of the marginalised one-dimensional posterior probabilities.

\begin{figure}
\begin{center}
\includegraphics[width=3in]{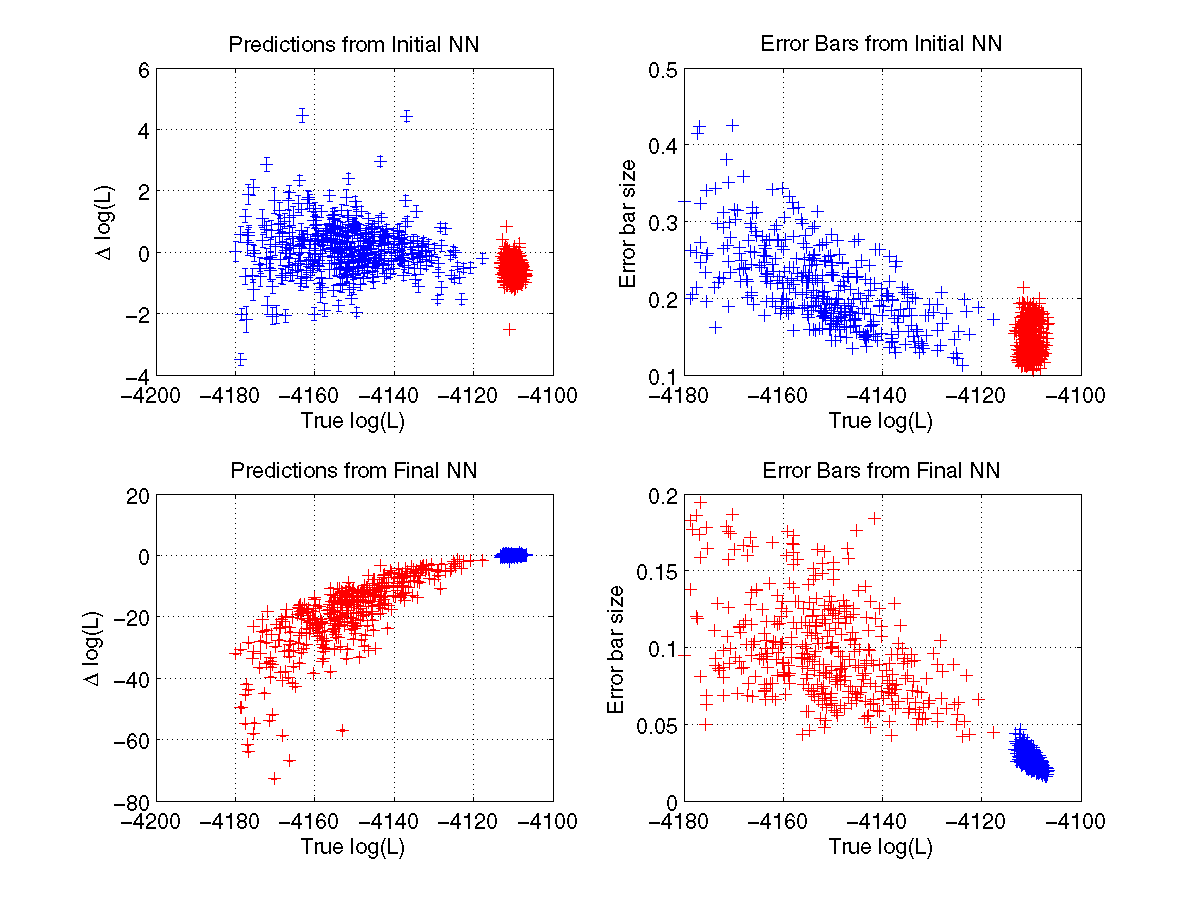}
\caption{Predictions with uncertainty error bars for NNs saved by BAMBI when analysing the non-flat model using the complete data set. The left-hand side shows predictions with errors for the two NNs on their own and the other's validation data sets. Each network's own points are in blue +s, the other NN's points are in red crosses. Many error bars are too small to be seen. The right-hand side, using the same colour and label scheme, shows the magnitudes of the error bars from each NN on the predictions.}
\label{fig:NNerrors}
\end{center}
\end{figure}

In Figure~\ref{fig:NNerrors} we show predictions from the two trained NNs on the two sets of validation data points in the case of the non-flat model using the complete data set. In the left-hand column, we can see that the first NN trained is able to make reasonable predictions on its own validation data as well as on the second set of points, in red crosses, from the second NN's validation data. The final NN is able to make more precise predictions on its own data set than the initial NN, but is unable to make accurate predictions on the first NN's data points. The right-hand column shows the error bar sizes for each of the points shown. For both NNs, the errors decrease with increasing log-likelihood. The final NN has significantly lower uncertainty on predictions for its own validation data, which enables us to set the threshold for when we can trust its prediction.
The cases for the other three sets of cosmological models and data sets are very similar to this one. This demonstrates the need to use the uncertainty error measurement in determining which NN prediction to use, if any. Always using the final NN would produce poor predictions away from the peak and the initial NN does not have sufficient precision near the peak to properly measure the best-fit cosmological parameters. But by choosing which NN's prediction to accept, as we have shown, we can quickly and accurately reproduce the likelihood surface for sampling. Furthermore, if one were interested only in performing a re-analysis about the peak, then one could use just the final NN, thereby omitting the calculational overhead associated with choosing the appropriate network.

\begin{figure}
\begin{center}
\includegraphics[width=3in]{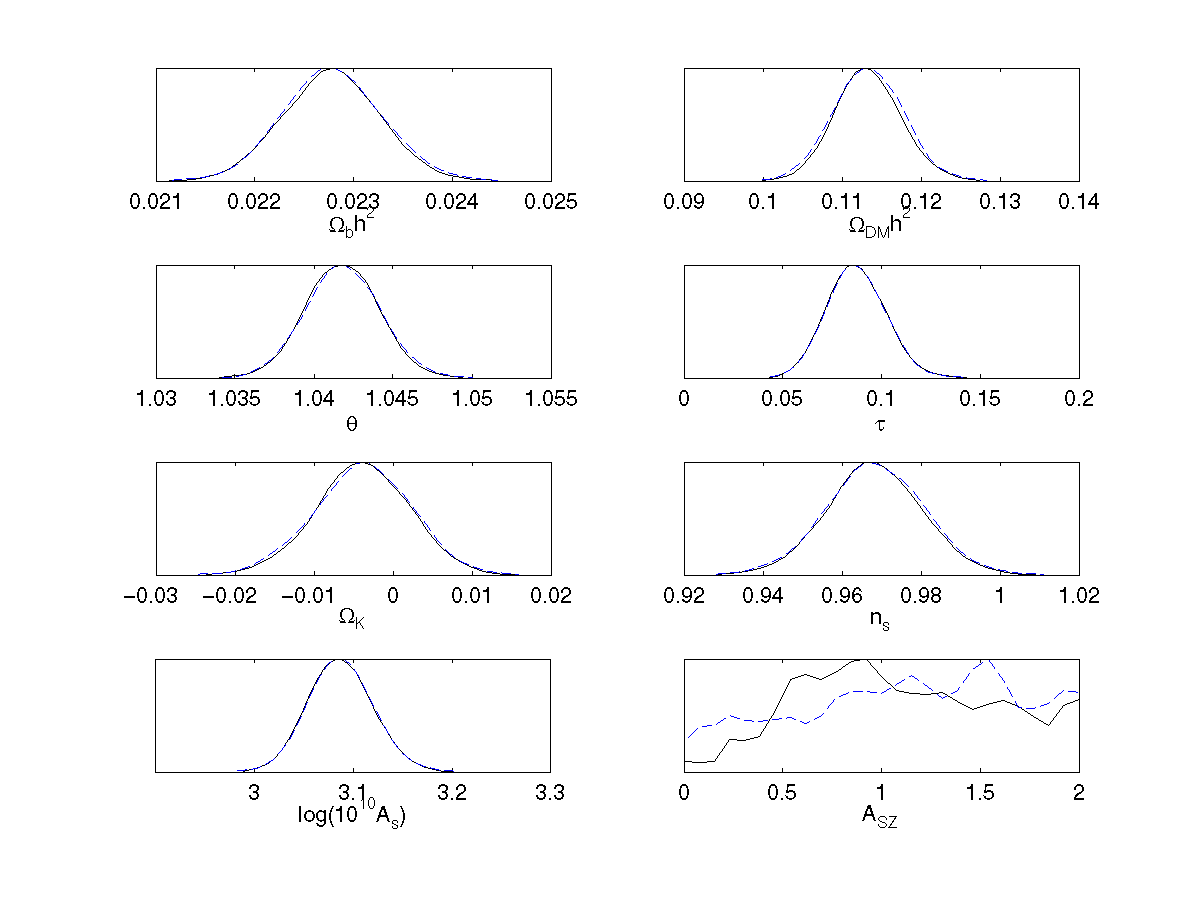}
\caption{Marginalised 1D posteriors for the non-flat model ($\Lambda$CDM$+\Omega_K$) using the complete data set. BAMBI's initial run is in solid black, the follow-up analysis in dashed blue.}
\label{fig:fastpostcomp1}
\end{center}
\end{figure}

\begin{figure}
\begin{center}
\includegraphics[width=3in]{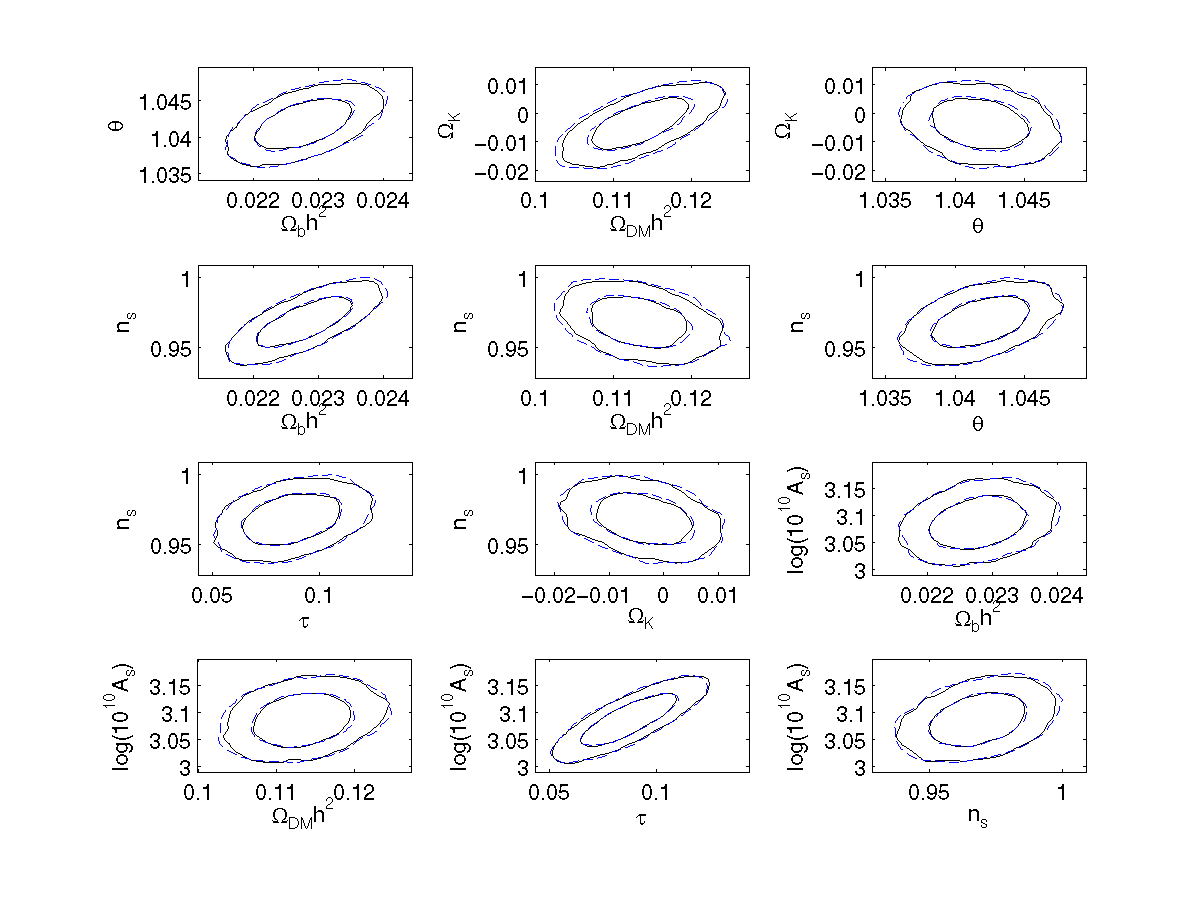}
\caption{Marginalised 2D posteriors for the non-flat model ($\Lambda$CDM$+\Omega_K$) using the complete data set. The $12$ most correlated pairs are shown. BAMBI's initial run is in solid black, the follow-up analysis in dashed blue. Inner and outer contours represent $68\%$ and $95\%$ confidence levels, respectively.}
\label{fig:fastpostcomp2}
\end{center}
\end{figure}

For this same case, we plot the one- and two-dimensional marginalised posterior probabilities in Figures~\ref{fig:fastpostcomp1} and~\ref{fig:fastpostcomp2}, respectively. Although the priors do not cover exactly the same ranges, we expect very similar posterior distributions since the priors are sufficiently wide as to encompass nearly all of the posterior probability. We see this very close agreement in all cases.

Calculating the uncertainty error requires calculating approximate inverse Hessian-vector products which slow down the process. We sacrifice a large factor of speed increase in order to maintain the robustness of our predictions. Using the same method as before, we computed the time per likelihood calculation for the initial BAMBI run as well as the follow up; these are compared in Table~\ref{tab:BAMBIfollow-up}. We can see that in addition to the initial speed-up obtained with BAMBI, this follow-up analysis obtains an even larger speed-up in time per likelihood calculation. This speed-up is especially large for the non-flat model, where CAMB takes longer to compute the CMB spectra. The speed factor also increases when using the complete data set, as the original likelihood calculation takes longer than for the CMB-only data set; NN predictions take equal time regardless of the data set.
\begin{table}
\begin{center}
\begin{tabular}{ccccc}
Model & Data set & Initial & Follow-up & Speed \\ 
 & & $t_{\mathcal{L}}$ (s) & $t_{\mathcal{L}}$ (s) & factor \\ \hline \hline
$\Lambda$CDM & CMB only & $1.635$ & $0.393$ & $4.16$ \\
$\Lambda$CDM & all & $2.356$ & $0.449$ & $5.25$ \\
$\Lambda$CDM$+\Omega_K$ & CMB only & $9.520$ & $0.341$ & $27.9$ \\
$\Lambda$CDM$+\Omega_K$ & all & $8.640$ & $0.170$ & $50.8$ \\
\end{tabular}
\caption{Time per likelihood evaluation, factor of speed increase from BAMBI's initial run to a follow-up analysis.}
\label{tab:BAMBIfollow-up}
\end{center}
\end{table}

One possible way to avoid the computational cost of computing error bars on the predictions is that suggested by~\cite{NNprederror}. One can take the NN training data and add Gaussian noise and train a new NN, using the old weights as a starting point. Performing many realisations of this will quickly provide multiple NNs whose average prediction will be a good fit to the original data and whose variance from this mean will measure the error in the prediction. This will reduce the time needed to compute an error bar since multiple NN predictions are faster than a single inverse Hessian-vector product. Investigation of this technique will be explored in a future work.

\section{Summary and Conclusions}\label{sec:conclusion}
We have introduced and demonstrated a new algorithm for rapid Bayesian data analysis. The Blind Accelerated Multimodal Bayesian Inference algorithm combines the sampling efficiency of {\sc MultiNest} with the predictive power of artificial neural networks to reduce significantly the running time for computationally expensive problems.

The first applications we demonstrated are toy examples that demonstrate the ability of the NN to learn complicated likelihood surfaces and produce accurate evidences and posterior probability distributions. The eggbox, Gaussian shells, and Rosenbrock functions each present difficulties for Monte Carlo sampling as well as for the training of a NN. With the use of enough hidden-layer nodes and training points, we have demonstrated that a NN can learn to accurately predict log-likelihood function values.

We then apply BAMBI to the problem of cosmological parameter estimation and model selection. We performed this using flat and non-flat cosmological models and incorporating only CMB data and using a more extensive data set. In all cases, the NN is able to learn the likelihood function to sufficient accuracy after training on early nested samples and then predict values thereafter. By calculating a significant fraction of the likelihood values with the NN instead of the full function, we are able to reduce the running time by a factor of $1.19$ to $1.42$. This is in comparison to use of {\sc MultiNest} only, which already provides significant speed-ups in comparison to traditional MCMC methods~\citep[see][]{MultiNest2}.

Through all of these examples we have shown the capability of BAMBI to increase the speed at which Bayesian inference can be done. This is a fully general method and one need only change the settings for {\sc MultiNest} and the network training in order to apply it to different likelihood functions. For computationally expensive likelihood functions, the network training takes less time than is required to sample enough training points and sampling a point using the network is extremely rapid as it is a simple analytic function. Therefore, the main computational expense of BAMBI is calculating training points while the sampling evolves until the network is able to reproduce the likelihood accurately enough. With the trained NN, we can now perform additional analyses using the same likelihood function but different priors and save large amounts of time in sampling points with the original likelihood and in training a NN. Follow-up analyses using already trained NNs provides much larger speed increases, with factors of $4$ to $50$ obtained for cosmological parameter estimation relative to BAMBI speeds. The limiting factor in these runs is the calculation of the error of predictions, which is a flat cost based on the size of the NN and data set, regardless of the original likelihood function.

The NNs trained by BAMBI for cosmology cover a larger range of log-likelihoods than the one trained for {\sc CosmoNet}. This allows us to use a wider range of priors for subsequent analysis and not be limited to the four-sigma region around the maximum likelihood point. By setting the tolerance for BAMBI's NNs to a larger value, fewer NNs with larger likelihood ranges can be trained, albeit with larger errors on the predictions. Allowing for larger priors requires us to test the validity of our NNs' approximations, which ends up slowing the overall analysis.

Since BAMBI uses a NN to calculate the likelihood at later times in the analysis where we typically also suffer from lower sampling efficiency (harder to find a new point with higher likelihood than most recent point removed), we are more easily able to implement Hamiltonian Monte Carlo~\citep{HamiltonianMC} for finding a proposed sample. This method uses gradient information to make better proposals for the next point that should be `sampled'. Calculating the gradient is usually a difficult task, but with the NN approximation they are very fast and simple. This improvement will be investigated in future work.

As larger data sets and more complicated models are used in cosmology, particle physics, and other fields, the computational cost of Bayesian inference will increase. The BAMBI algorithm can, without any pre-processing, significantly reduce the required running time for these inference problems. In addition to providing accurate posterior probability distributions and evidence calculations, the user also obtains a NN trained to produce likelihood values near the peak(s) of the distribution that can be used in even more rapid follow-up analysis.

\section*{Acknowledgments}
PG is funded by the Gates Cambridge Trust and the Cambridge Philosophical Society; FF is supported by a Trinity Hall College research fellowship. The authors would like to thank Michael Bridges for useful discussions, Jonathan Zwart for inspiration in naming the algorithm, and Natallia Karpenko for helping to edit the paper. The colour scheme for Figures~\ref{fig:eggboxlike}, \ref{fig:gausslike}, and~\ref{fig:rosenbrocklike} was adapted to MATLAB from~\cite{CubeHelix}. This work was performed on COSMOS VIII, an SGI Altix UV1000 supercomputer, funded by SGI/Intel, HEFCE and PPARC and the authors thank Andrey Kaliazin for assistance. The work also utilised the Darwin Supercomputer of the University of Cambridge High Performance Computing Service (\texttt{http://www.hpc.cam.ac.uk/}), provided by Dell Inc. using Strategic Research Infrastructure Funding from the Higher Education Funding Council for England and the authors would like to thank Dr. Stuart Rankin for computational assistance.

\setlength{\labelwidth}{0pt}


\label{lastpage}

\end{document}